\documentclass[twocolumn]{aastex62}

\usepackage{natbib}
\usepackage{amsmath}

\hypersetup{linkcolor=red,citecolor=blue,filecolor=cyan,urlcolor=magenta}
\newcommand{\BB}{{\cal B}}

\graphicspath{{./}{figures/}}

\received{December 15, 2019}
\revised{May 2, 2020}
\accepted{May 28, 2020}
\shorttitle{Effect of Electromagnetic Interaction on Galactic Center Flare Components
}
\shortauthors{Tursunov et al.}
\begin{document}

\title{Effect of Electromagnetic Interaction on Galactic Center Flare Components}

\correspondingauthor{Arman Tursunov}
\email{arman.tursunov@physics.slu.cz }

\author[0000-0001-5845-5487]{Arman Tursunov}
\affil{I. Physikalisches Institut der Universit\"at zu K\"oln, Z\"ulpicher Strasse 77, D-50937 K\"oln, Germany}
\affil{Research Centre for Theoretical Physics and Astrophysics, Institute of Physics, \\Silesian University in Opava, Bezru\v{c}ovo n\'{a}m.13, CZ-74601 Opava, Czech Republic}

\author[0000-0001-6450-1187]{Michal Zaja\v{c}ek}
\affiliation{Center for Theoretical Physics, Polish Academy of Sciences, Al. Lotnikow 32/46, 02-668 Warsaw, Poland}
\affiliation{Max-Planck-Institut f\"ur Radioastronomie (MPIfR), Auf dem H\"ugel 69, D-53121 Bonn, Germany}
\affiliation{I. Physikalisches Institut der Universit\"at zu K\"oln, Z\"ulpicher Strasse 77, D-50937 K\"oln, Germany}

\author[0000-0001-6049-3132]{Andreas Eckart}
\affiliation{I. Physikalisches Institut der Universit\"at zu K\"oln, Z\"ulpicher Strasse 77, D-50937 K\"oln, Germany}
\affiliation{Max-Planck-Institut f\"ur Radioastronomie (MPIfR), Auf dem H\"ugel 69, D-53121 Bonn, Germany}

\author[0000-0002-4900-5537]{Martin Kolo{\v s}}
\affil{Research Centre for Theoretical Physics and Astrophysics, Institute of Physics, \\Silesian University in Opava, Bezru\v{c}ovo n\'{a}m.13, CZ-74601 Opava, Czech Republic}

\author[0000-0001-9240-6734]{Silke Britzen}
\affiliation{Max-Planck-Institut f\"ur Radioastronomie (MPIfR), Auf dem H\"ugel 69, D-53121 Bonn, Germany}

\author[0000-0003-2178-3588]{Zden{\v e}k Stuchl\'ik}
\affil{Research Centre for Theoretical Physics and Astrophysics, Institute of Physics, \\Silesian University in Opava, Bezru\v{c}ovo n\'{a}m.13, CZ-74601 Opava, Czech Republic}

\author[0000-0001-5848-4333]{Bozena Czerny}
\affiliation{Center for Theoretical Physics, Polish Academy of Sciences, Al. Lotnikow 32/46, 02-668 Warsaw, Poland}

\author[0000-0002-5760-0459]{Vladim\'ir Karas}
\affiliation{Astronomical Institute, Czech Academy of Sciences, Bo\v{c}n\'{\i}~II 1401, CZ-14131~Prague, Czech Republic}



\begin{abstract}
Recently, near-infrared  GRAVITY@ESO observations at $2.2\,\mu{\rm m}$ have announced the detection of three bright ``flares" in the vicinity of the Galactic center supermassive black hole (SMBH) that exhibited orbital motion at a { distance  of} about $6 - 11$ gravitational radii from an $\sim 4\times 10^6\, M_{\odot}$ black hole.  There are indications of the presence of a large-scale, organized component of the  magnetic field at the Galactic center. Electromagnetic effects on the flare dynamics were previously not taken into account despite the relativistic motion of a plasma in magnetic field leading to the charge separation and nonnegligible net charge density in the plasma. Applying various approaches, we find the net charge number density of the flare components of the order of $10^{-3} - 10^{-4}$ cm$^{-3}$, while the particles' total number density is of the order of $10^{6} - 10^{8}$ cm$^{-3}$. However, even such a tiny excess of charged particles in the quasi-neutral plasma can { significantly} affect the dynamics of flare components, which can then lead to the { degeneracy} in the measurements of spin of the SMBH. Analyzing the dynamics of recent flares in the case of the rapidly rotating black hole, we also constrain the inclination angle between the magnetic field and spin axis to $\alpha < 50^{\circ}$, as for larger angles, the motion of the hot spot is strongly chaotic.  
\end{abstract}



\section{Introduction} \label{sec:intro}

The compact radio source Sgr~A*  at the Galactic center \citep{1974ApJ...194..265B} associated with the supermassive black hole (SMBH) and the dynamical center of our Galaxy is a highly variable source across all wavelengths \citep{2005bhcm.book.....E,2007gsbh.book.....M,2010RvMP...82.3121G,2017FoPh...47..553E,2019arXiv190106507K}. { {Given its mass of $\sim 4\times 10^6$ Solar masses inferred from stellar dynamics \citep[see, e.g.][]{2013ApJ...779L...6D,2016ApJ...830...17B,2017ApJ...837...30G,2017ApJ...845...22P}, as well as from bright X-ray flares \citep{2017MNRAS.472.4422K,2019arXiv190106520K}, it has been considered as one of the best candidates for an SMBH. From the early theoretical predictions \citep{1971MNRAS.152..461L}, there have been several key experiments that very precisely measured the gravitational redshift of the fast-moving S2 star during its pericenter passage in 2018 May \citep{2018A&A...615L..15G}, as well as its Schwarzschild precession of $\delta \phi\sim 12'$ per orbital period \citep{2020arXiv200407187G}, which is fully consistent with the general relativistic predictions. In addition, the very long baseline interferometry (VLBI) observations of Sgr~A* at 1.3 mm \citep{2008Natur.455...78D} indicate a source structure at event-horizon scales, which is promising for the detection of the shadow \citep{1973blho.conf..215B,2000ApJ...528L..13F} in a similar way as was performed for M87 \citep{2019ApJ...875L...1E}. Hence, Sgr~A* can be considered as an SMBH, with little space for alternative scenarios such as a boson star, gravastar, or fermion ball \citep{2017FoPh...47..553E,2019JPhCS1258a2031Z}.}}

The source structure of Sgr~A* and its temporal changes were resolved out on event-horizon scales by the VLBI technique at $1.3\,{\rm mm}$ \citep{2008Natur.455...78D,2011ApJ...727L..36F,2015Sci...350.1242J,2018ApJ...859...60L}, which showed that the bulk of emission of Sgr~A* may not be centered at the black hole itself. The VLBI study by \citet{2015Sci...350.1242J} inferred from the linearly polarized emission at $1.3\,{\rm mm}$ that a partially ordered magnetic field is present on the scale of 6 to 8 Schwarzschild radii. They also detected an intrahour variability time-scale associated with this field. These findings are consistent with the recent GRAVITY@ESO\footnote{An NIR, beam-combining interferometry instrument operating in the $K_{\rm s}$-band continuum that is capable of high-resolution imaging (resolution of $3\,{\rm mas}$) and astrometry (resolution $\sim 20-70\,{\mu as}$).} discovery of continuous positional and polarization offsets of emission centroids during high states of Sgr~A* activity, so-called ``flares'', in the near-infrared (NIR) $K_{\rm s}$-band $(2.2\,{\rm \mu m})$ continuum emission \citep{2018A&A...618L..10G}. The linear polarization angle turns around continuously with {a} period comparable to the orbital motion of the emission centroid (hereafter called the \textit{hot spot}), $P_{\rm hs}=45(\pm 15)\,{\rm min}$, which implies { an} ordered poloidal magnetic field, i.e. perpendicular to the orbital plane, while for the toroidal geometry, one expects two polarization loops per orbital period \citep{2001ApJ...555L..83B,2016MNRAS.462..115D}. { The presence of a dynamically significant magnetic field in the accretion zone close to the innermost stable circular orbit (ISCO) is also consistent with the magnetic field strength of $B\geq 8\,{\rm mG}$ at the larger projected distance of $R\sim 0.12\,{\rm pc}$, as inferred from the Faraday rotation measurements of the magnetar J1745-2900 \citep{2013Natur.501..391E}. As Sgr~A* accretes from the magnetized plasma at larger scales, the magnetic field and the plasma density are expected to further increase inward.}

\subsection{Broadband spectral characteristics of Sgr~A*} \label{subsec:broadband}
The present activity of Sgr~A* is very low, and in comparison with active galactic nuclei (AGN), it can be generally characterized as extremely low-luminous \citep{2010RvMP...82.3121G,2017FoPh...47..553E}, which stems from the comparison of its theoretical Eddington limit,
\begin{equation}
    L_{\rm Edd}=5 \times 10^{44}\left(\frac{M}{4\times 10^6\,M_{\odot}} \right)\,{\rm erg\,s^{-1}}\,,
    \label{eq_eddington_limit}
\end{equation}
and its 8 orders-of-magnitude smaller bolometric luminosity of $\sim 10^{36}\,{\rm erg\,s^{-1}}$ inferred from observations and explained by radiatively inefficient accretion flow models \citep[RIAFs;][]{1998ApJ...492..554N,1999MNRAS.303L...1B}. The mass of Sgr~A* in Eq.~\eqref{eq_eddington_limit} is scaled to the value of $\sim 4\times 10^6\,M_{\odot}$ derived from the most recent S2 star observations by the \citet{2018A&A...615L..15G} \citep[see also][ for comparison]{2016ApJ...830...17B,2017ApJ...845...22P,2017ApJ...837...30G}, which corresponds to the gravitational radius of $R_{\rm g}=GM/c^2=5.9 \times 10^{11}\,{\rm cm}\sim 10^{12}\,{\rm cm}$ that we apply in the further analysis.

The object Sgr~A* is surrounded by $\sim 200$ massive He I emission-line stars of spectral type OB, and it is thought to capture their wind material with an estimated rate of $\dot{M}_{\rm B} \approx 10^{-5}\,M_{\odot}\,{\rm yr^{-1}}$ at a Bondi radius of $r_{\rm B}=4''(T_{\rm a}/10^7\,{\rm K})\approx 0.16\,{\rm pc}=8.1\times 10^5\,R_{\rm g}$ \citep{2003ApJ...591..891B,2010ApJ...716..504S,2019MNRAS.482L.123R}, where the gravitational pull of the SMBH prevails over that of the thermal gas pressure with temperature $T_{\rm a}$. From the submillimeter Faraday rotation measurements within the inner $r\lesssim 200\,R_{\rm g}$ it was inferred that Sgr~A* accretes at least 2 orders of magnitude less than the Bondi rate, $\dot{M}_{\rm acc}\approx 2\times 10^{-7}-2\times 10^{-9}\,M_{\odot}\,{\rm yr^{-1}}$ \citep{2007ApJ...654L..57M}; hence, most of the material captured at the Bondi radius is expelled and leaves the system 
 as an outflow, which is also consistent with the RIAF solutions with a density profile in the power-law form $n(r)\propto r^{-p}$, where $p \lesssim 1$ \citep{2013Sci...341..981W}. The density profile of the hot RIAF flow is flatter than the density profile of the stationary spherical Bondi accretion, for which $n(r)\propto r^{-3/2}$. This flattening is caused by the presence of outflows \citep{2012ApJ...761..130Y,2013Sci...341..981W}. The inflow-outflow RIAF models (disk-jet/wind or advection-dominated accretion flow (ADAF) and jet -- jet-ADAF) can inhibit the accretion rate on smaller spatial scales by the transport of energy released during accretion to larger radii \citep{2002A&A...383..854Y,2010ApJ...716..504S,2014A&A...570A...7M,2015ApJ...799....1C,2017MNRAS.467.3604R}, which reduces the accretion rate to $\lesssim 1\%$ of the Bondi rate, and the jet-ADAF models can generally capture the main features of the Sgr~A* broadband spectrum.
The extremely low luminosity of Sgr~A* is thus best explained by the combination of a low accretion rate $\dot{M}_{\rm acc}$ and very low radiative efficiency of the accretion flow $\eta_{\rm acc} \approx 5\times 10^{-6}$ \citep{2014ARA&A..52..529Y}, which is 4 orders of magnitude below the standard $10\%$ efficiency applicable to luminous AGN with significantly higher accretion rates.  

In the radio/millimeter domain, the flux density generally increases with frequency with a rising spectral index from $\alpha=0.1-0.4$ to $0.76$ at 2-3 mm\footnote{Using the notation $S_{\nu}\propto \nu^{\alpha}$, where $S_{\nu}$ is the monochromatic flux density in Janskys ($1~ {\rm Jy}=10^{-23}\,{\rm erg\,s^{-1}\,Hz^{-1}\,cm^{-2}}$), $\nu$ is the frequency in Hertz (Hz), and $\alpha$ is the spectral index. In the radio domain, the spectral index $\alpha<0$ is referred to as steep (optically thin synchrotron emission), $\alpha>0$ is referred to as inverted (self-absorbed, optically thick synchrotron emission), and $\alpha\sim 0$ stands for a flat spectral profile.} and with a clear peak or bump close to $1\,{\rm mm}$, which is referred to as the submillimeter bump \citep[e.g., ][]{1998ApJ...499..731F,2010ApJ...717.1092D,2015ApJ...802...69B}.   The submillimeter bump is produced by the optically thick synchrotron emission that originates from relativistic, thermal electrons (with a Lorentz factor of $\gamma_{\rm e} \sim 10$) in the innermost portions of the hot, thick ADAF \citep{1995Natur.374..623N,1998ApJ...492..554N,2003ApJ...598..301Y}. It marks the transition from the optically thick emission at lower frequencies to the optically thin emission at higher frequencies \citep{1995A&A...297...83Z,1997ApJ...490L..77S,1998ApJ...499..731F}. Below $1\,{\rm mm}$, the medium gets optically thin and the flux density gradually drops all the way { to} X-ray wavelengths, where the quiescent counterpart of Sgr~A* was detected with the unabsorbed 2-10 keV luminosity of $L_{\rm x}\approx 2 \times 10^{33}\,{\rm erg\,s^{-1}}$ { produced by thermal bremsstrahlung from cooler electrons at larger distances close to the Bondi radius} \citep{2003ApJ...591..891B,2010ApJ...716..504S}, with no detected quiescent counterpart in the infrared (IR) domain \citep[see, however, the upper limits on the far-IR flux density based on the detected variability by][]{2018ApJ...862..129V}. { Thanks to high-sensitivity GRAVITY observations in the $K_{\rm s}$ band ($2.2\,{\rm \mu m}$) \citep{2020arXiv200407185T}, it was possible to detect a turnover in the flux density distribution of NIR flares with a median value of $(1.1 \pm 0.3)\,{\rm mJy}$. The flux density distribution in the NIR domain was found to have two states: the bulk of the emission can be described by a log-normal distribution with a median around $1.1\,{\rm mJy}$, and on top of this quiescent emission, there are the sporadic flares at higher flux densities with a single power-law distribution. A single power-law or log-normal distribution cannot describe the flux density distribution as a whole.} 

While in the radio/millimeter domain, Sgr~A* is mildly variable, it exhibits order-of-magnitude nonthermal high states or flares in the IR and X-ray domain a few times per day on a timescale of $\sim 1$ hr \citep{2001Natur.413...45B,2003Natur.425..934G,2004ApJ...601L.159G,2006A&A...455....1E,2010A&A...510A...3Z,2012A&A...537A..52E,2012ApJS..203...18W,2017MNRAS.472.4422K,2018ApJ...863...15W,2018A&A...618L..10G}, with the X-ray flares always being simultaneously associated with IR flares but not vice versa. The IR flares are linearly polarized with a polarization degree of $20\%\pm 15\%$ and a rather stable polarization angle of $13^{\circ} \pm 15^{\circ}$ \citep{2015A&A...576A..20S}, which likely reflects the overall stability of the disk-jet system of Sgr~A*.

\begin{figure*}[tbh]
    \centering
    \includegraphics[width=1.3\columnwidth
    ]{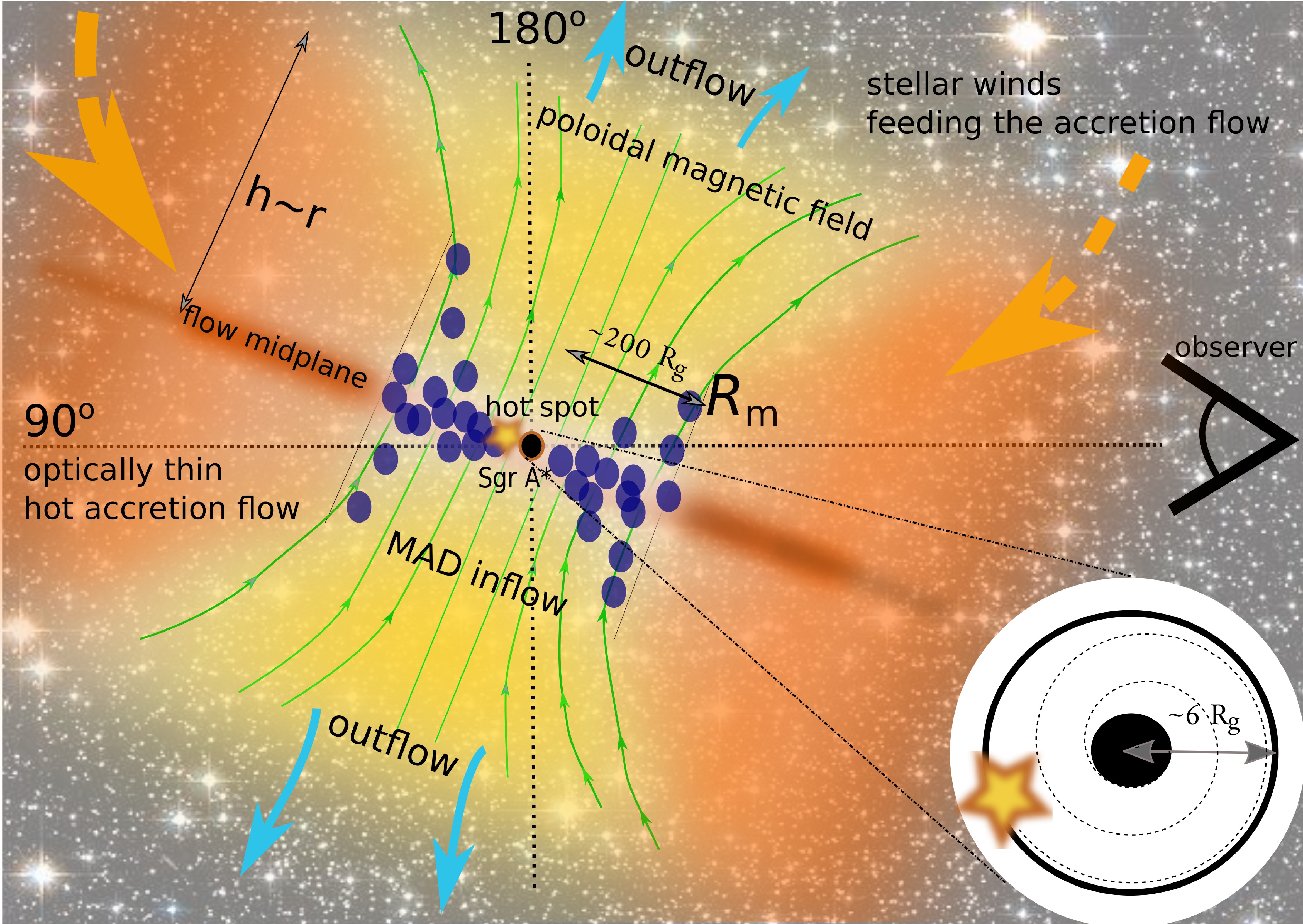}
    \caption{ Illustration of the hot accretion flow dominantly supplied by stellar winds with the inner MAD \citep{2003PASJ...55L..69N} part due to the accumulation of the dominant poloidal magnetic field. The transition from the hot, thick flow into clumpy flow occurs at the magnetospheric radius $R_{\rm m}$, which is expected to be located at $\sim 200 R_{\rm g}$, see Eq.~\ref{eq_magnetospheric_radius}. The MAD consists of magnetically confined blobs that diffuse inward through the poloidal magnetic field through the processes of magnetic reconnection and magnetic interchanges. At the same time, the released energy heats up the surrounding gas, which forms a hot and diluted corona and powers outflows. The figure inset to the right captures the innermost part of the accretion flow close to the ISCO (located at $6\,R_{\rm g}$ for a nonrotating black hole), where the hot spot orbits Sgr~A* for a large fraction of its orbital period and then presumably plunges toward the event horizon. Its emission is then detected at the Earth in the NIR domain as a transient flare, whose flux density is modulated by the Doppler boosting and the gravitational lensing \citep[see, e.g.][]{2017FoPh...47..553E}. 
		The stellar field in the background is added to show that stellar winds of massive OB stars supply a large portion of the hot thick flow whose outer radius is approximately at the Bondi radius at $\sim 0.16$pc. The dense stellar field originally belongs to the globular cluster NGC 288 imaged by the Hubble Space Telescope's Wide Field Channel of the Advanced Camera for Surveys (credit: ESA/Hubble \& NASA). }
    \label{fig_illustration_MAD}
\end{figure*}

%%%%%%%%%%%%%% INTRO: FLARES AND HOT SPOT CHARGE %%%%%%%%%%%%
\subsection{Flare--Hot spot connection} \label{subsec: flare-hot spot}
The $K_{\rm s}$-band observations by GRAVITY@ESO \citep{2018A&A...618L..10G} brought the first direct evidence that flares are associated with { the orbiting luminous mass} or hot spots.
The GRAVITY observations of hot spots close to the ISCO of Sgr~A* have enabled the fitting of their orbital periods as well as orbital radii with the equatorial circular orbits of neutral test particles around rotating Kerr black hole of mass $\sim 4$ million $M_\odot$. 

The origin and nature of flares/hot spots still remains unclear. Despite many suggestions, including their connection to the tidal disruption of asteroids \citep{2012MNRAS.421.1315Z}, they are most likely connected to dynamical changes in the hot, magnetized accretion flow. As transient phenomena, they could originate from  magnetohydrodynamic instabilities or magnetic reconnection events, as is the case of X-ray flares on the Sun \citep{2009MNRAS.395.2183Y} or has been discussed for M87 \citep{2017A&A...601A..52B}. The model of ejected plasmoids during reconnection events is supported by their statistical properties; namely, the count rate  versus flux density distribution can be fitted with the power law in the X-ray, IR, submillimeter and radio domain, $\mathrm{d}N/\mathrm{d}E\propto E^{-\alpha}\mathrm{d}E$, which is consistent with the self-organized criticality phenomena of spatial dimension $S=3$ \citep{2012ApJS..203...18W,2015ApJ...810...19L,2017A&A...601A..80S,2018ApJ...863...15W}.   

 There are other mechanisms that lead to the power-law distribution of plasmoid properties. Namely, \citet{2010PhRvL.105w5002U} showed analytically that the tearing (plasmoid) instability during magnetic reconnection leads to the power-law distribution of the magnetic flux $\psi$ in plasmoids in high Lundquist number current sheets, $f(\psi)\propto \psi^{-2}$, with an exponential decay in the tail of the distribution \citep{2010PhPl...17a0702F}, while \citet{2012PhRvL.109z5002H}, using direct numerical simulations,  showed that the slope is smaller, $f(\psi)\propto \psi^{-1}$. In addition, magnetohydrodynamic (MHD) turbulence driven by the magnetorotational instability can lead to a power-law distribution of dissipative events \citep[see, e.g.][]{2015PhRvL.114f5002Z}. \citet{2015PhRvL.114f5002Z} found the probability distribution of the dissipated energy with an index of $\alpha=1.75 \pm 0.10$ and the distribution of energy dissipation rates with a slope of $\alpha \sim 2$, which is two times less than the NIR, submillimeter, and radio distribution of flare flux densities, for which $\alpha=4$ \citep[using the notation $p(x)\propto x^{-\alpha}$, see][]{2012ApJS..203...18W,2017A&A...601A..80S}, but consistent with the energy distribution of X-ray flares, for which a slope of $\alpha=1.65 \pm 0.17$ was recovered \citep{2015ApJ...810...19L}. This is strikingly similar to the solar flare energy distribution with an index of $\alpha \sim 1.8$ \citep{1991SoPh..133..357H}. \citet{2013ApJ...774...42N} found a slope of $1.9^{+0.4}_{-0.5}$ for the X-ray flare peak rate distribution and $1.5 \pm 0.2$ for the fluence distribution, which is also consistent with the finding of \citet{2015PhRvL.114f5002Z} that the energy (fluence) distribution is shallower than the peak rate distribution. For the future, it will be necessary to verify whether the analytical and numerical studies of MHD turbulence and instabilities are also applicable to the plasma in the strong gravity regime in 3D and whether the tearing and the magnetohydrodynamic turbulence can lead to the power-law flux distribution of the flares that are sampled over a longer period of time.

 To link the plasmoid model with the variability of Sgr~A*, a plasma blob or plasmon that cools down via the adiabatic expansion was applied to fit simultaneous multiwavelength flares. \citet{2008ApJ...682..361Y} found a time-lag of $110 \pm 17$ minutes between X-ray and submillimeter $850\,{\rm \mu m}$ flares and the time-lag of $\sim 20-30$ minutes between 7 and 13 mm flare peaks. The time delays are matched well by an initially optically thick cloud of synchrotron-emitting electrons that becomes optically thin towards consecutively lower frequencies as it expands and cools down adiabatically. The basic explanation of the delay between X-ray and submillimeter flare peak emissions is that at first, the emission is optically thick in the submillimeter domain, and later, it gets optically thin due to the blob expansion \citep{1966Natur.211.1131V}. Furthermore, the model can reproduce well the asymmetric profile of the light curves (faster rising, slower fading) and the linear polarization degree of $\sim 1\%$ at radio bands. The inferred parameters include the comoving expansion velocity of the blob, $v_{\rm exp}\sim 0.003-0.1c$, the magnetic field in the range $10-70\,{\rm G}$, and a particle spectral index of $\alpha=-1.5\pm 0.5$. With this expansion velocity, plasma cannot escape from Sgr~A*, unless a large bulk motion is present. These results have been confirmed independently by \citet{2010A&A...517A..46K}, \citet{2012A&A...537A..52E}, and \citet{2016MNRAS.458.2336B}. The NIR flares are explained via the synchrotron emission of localized, heated relativistic electrons, and the X-ray flares that are simultaneous with NIR flares are explained via three processes: (i) synchrotron emission of the same population of electrons as for NIR flares with a power-law distribution of energies, (ii) the Compton upscattering of submillimeter seed photons by NIR-emitting electrons to X-ray energies (external inverse Compton), and (iii) the NIR synchrotron photons upscattered to X-ray energies by the same population of heated electrons that produce NIR flares, which is referred to as the synchrotron self-Compton (SSC) mechanism \citep[see][for a review]{2010RvMP...82.3121G}.

Another model to explain the hot spot phenomenon and the associated short-term variability in the Galactic center would be vortices and magnetic field flux tubes that could be sites of dissipation and collimated radiation \citep{1992Natur.356...41A}. Vortices as coherent structures are typical for any rotating fluid; hence, this scenario could also be applicable to the Galactic center hot flow.

 The simplest explanation of the hot spot nature is given by the discrete origin of the NIR flares. The single-state stochastic nature of Sgr~A* variability \citep{2014ApJ...791...24M} suggests a clumpy mode of accretion. The hot spot would then correspond to confined islands or blobs of heated plasma that descend down the potential well toward the event horizon. This view is consistent with the class of magnetically arrested accretion flow models \citep[MADs; ][]{1974Ap&SS..28...45B,2003PASJ...55L..69N,2008ApJ...677..317I}, in which the accretion flow becomes unstable due to the accumulation of the magnetic flux and fragments into magnetically confined blobs below the magnetospheric radius,
\begin{align}
    R_{\rm m} & \sim  \frac{8\pi G M \rho}{B_{\rm pol}^2} =  \\ \,
              & =  181 \left(\frac{M}{4\times 10^6\,M_{\odot}}\right) \left(\frac{n_{\rm acc}}{10^6\,{\rm cm^{-3}}}\right) \left(\frac{B_{\rm pol}}{10\,{\rm G}}\right)^{-2}\, R_{\rm g}\,,\label{eq_magnetospheric_radius}
\end{align}
where the mass density is given by $\rho_{\rm acc}=\mu m_{\rm H}n_{\rm acc}$ ($\mu\sim 0.5$ for a fully ionized plasma), the poloidal magnetic field $B_{\rm pol}$ is scaled to $10\,{\rm G}$, and the number density is scaled to $10^6\,{\rm cm^{-3}}$, according to the values obtained from the synchrotron emission models of the flares \citep{2006ApJ...650..189Y,2008ApJ...682..361Y,2012A&A...537A..52E}. For such an accretion flow model, see also Fig.~\ref{fig_illustration_MAD} for an illustration where an initially hot, diluted, and thick axisymmetric flow fragments into clumps below $R_{\rm m}$. The clumpy structure of the flow is expected to dominate on a length scale of 100 $R_{\rm g}$ with a certain filling factor. The clumpy flow proceeds inwards diffusively through the poloidal magnetic field with the help of magnetic interchanges and reconnection events and the radial velocity is less than the freefall velocity \citep{2003PASJ...55L..69N}. The volume filling factor of the hot spots can be estimated $f_{\rm V}=V_{\rm hs}/V_{\rm acc}=n_{\rm acc}/n_{\rm hs}$, where $V_{\rm hs}$ is the volume of the hot spots, $V_{\rm acc}$ is the total volume of the accretion flow, $n_{\rm acc}$ is the mean number density of the accretion flow, and $n_{\rm hs}$ is the hot spot number density. The hot spot number density is thus expected to be larger than the mean density of the accretion flow, $n_{\rm hs}=n_{\rm acc}/f_{\rm V}$, depending clearly on $f_{\rm V}$. Typically, only one hot spot is present at the ISCO, according to NIR images and time series \citep[two to three events per day,][]{2018A&A...618L..10G}; therefore, within the ISCO volume, $f_{\rm V}\sim (R_{\rm hs}/r_{\rm ISCO})^3$, which for $R_{\rm hs}\sim 1 R_{\rm g}$ and a nonrotating black hole leads to $n_{\rm hs}\sim 216\, n_{\rm acc}$. In this setup, the hot spot would clearly be an overdense blob with respect to the background medium. 

\subsection{Hot spot inside Sgr A* magnetosphere}
The relativistic motion of a plasma around a black hole in the presence of the ordered magnetic field component (orthogonal to the orbital plane) necessarily leads to the charge separation and the consequent growth of its net charge density. This charge increases due to the compensation of the electric field in the comoving frame induced by the motion of plasma in the external magnetic field. In the case where the rotating plasma is associated with a neutron star threaded by a magnetic field, the arising charge density is known as the Goldreich-Julian (GJ) charge density \citep{1969ApJ...157..869G}. An analogous argument can be used to obtain a net charge density of the plasma of the flare components moving around the black hole, as will be described in Section~\ref{sec:GJ-argument}.

On the other hand, a similar charging process occurs near the black hole due to frame-dragging effect of the twisting of the magnetic field lines. In this scenario, both the black hole and the magnetosphere possess a nonnegligible electric charge \citep{1975PhRvD..12.2959R}. \citet{1978PhRvD..17.1518D} derived three different regions of charge particle trapping based on the interplay of electric and magnetic fields in the black hole magnetosphere. A magnetic field near a rotating black hole also plays a crucial role in the collimation of charged particles and thus also in the precollimation of astrophysical jets \citep{1997GReGr..29.1011K}.

For a black hole in this scenario, the charging mechanism was introduced by \cite{1974PhRvD..10.1680W}. Such an induced charge of Sgr~A* has an upper limit of $Q_{\rm BH} \sim 10^{15}$C \citep{2018MNRAS.480.4408Z}, which is still quite weak to have a gravitational effect on the spacetime metric; however, its electrostatic counterpart is nonnegligible for the motion of charged matter. Similarly, in a nonrelativistic case, the net charge of a massive object arises due to its rotation in the magnetic field \citep{1973ApL....13..109R}.   Computing Maxwell equations inside the magnetosphere \citep[see, e.g.,][]{1975PhRvD..12.2959R}, one can find that the total charge of the black hole magnetosphere is equal to the charge of the black hole with the   opposite sign, i.e. $Q_{\rm mag} = - Q_{\rm BH}$. This condition holds for a large class of accretion models. 

One of the interesting outcomes of the charge separation process in a plasma surrounding a black hole (and the consequent growth of the net charge densities of both the black hole and the magnetosphere) is that the black hole may act as a pulsar \citep{PhysRevD.98.123002}. For the dynamics of the flare components, furthermore, this leads to the inclusion of additional "electrostatic" interaction between the black hole and the hot spots if the black hole charge is not screened effectively.

In the Galactic center, a partially ordered magnetic field in the vicinity of Sgr~A* has been estimated with the strength of $10-100$ G. The NIR observations of horizontal polarization loops with the timescales comparable to the orbital periods of the recently observed bright flares imply the prevalent orientation of magnetic field lines in a direction that is perpendicular to the orbital plane of the corresponding hot spots. This implies that the hot spots associated with flares orbiting at the relativistic orbits are expected to possess a net electric charge due to the charge separation in the Galactic center plasma.

In this paper, we focus on the possible interplay between gravitational and electromagnetic fields in the interpretation of the observational features of the flares. 
Orthogonal orbital orientation with respect to the magnetic field lines of the three most recent hot spots and the charge separation in plasma leads to the appearance of an external Lorentz force arising from interactions of the flare components with the magnetic field. Taking into account the error bars of the GRAVITY measurement arising mainly due to astrometric errors and incomplete orbital coverage, we put limits on the strength of the Lorentz force, electric charges of hot spots, and net charge densities.

Below, we consider hot spot models with various charging mechanisms compared with observational data. It is worth noting that all obtained constraints on the charge values of hot spots have a comparable order of magnitude.

The paper is structured as follows. In Section~\ref{sec:parameters} we introduce the basic equations describing the model and provide estimates on the magnetic field, black hole charge, size, and mass of the flare components. In Section~\ref{sec:charge-limit} we study the charge separation in the plasma surrounding the black hole and estimate its net charge density. Applied to three recent flare components, we put tighter constraints on the hot spots charge based on their dynamics with inclusion of the electromagnetic interaction and the synchrotron radiation from hot spots. We also calculate the shifts of the ISCO caused by the interplay between the hot spot charge and external magnetic field and discuss the results. In Section~\ref{sec:sub35} we study the possible inclination of the orbital plane and magnetic field lines with respect to the orientation of the black hole's spin and put constraints on the inclination angle. We discuss the main results and their consequences in Section~\ref{sec:discussion} and give conclusions in Section~\ref{sec:conclusion}.

Throughout the paper, we use the space-like signature ($-,+,+,+$) and the system of units in which $c = 1$ and $G = 1$. However, for the expressions with an astrophysical application and estimates, we use the units with the gravitational constant and the speed of light. Greek indices are taken to run from 0 to 3; Latin indices are related to the space components of the corresponding equations.

%However, error bars of the measurement arising mainly due to astrometric errors and incomplete orbital coverage does not allow one to provide convincing conclusions on the consistency of the neutral hot spot model. Moreover, the mean values of periods of two flares (Jul 22 and May 27) appear to be slightly lower than the one predicted by simple neutral hot spot model.
% Orthogonal orbital orientation with respect to the magnetic field in all three flares and lower mean values of orbital periods may indicate the presence of an external force arising from interactions of flares with external magnetic field. In a simplified model of compact hot spot such an external force can be associated with the Lorentz force, given that the hot spot is slightly charged. 

\section{Model Setup and Hot Spot Parameters} \label{sec:parameters}

\subsection{ Model Assumptions}

 Our modeling approach is primarily motivated by NIR observations in the $K_{\rm s}$ band \citep[$2.2\,{\rm \mu m}$;][]{2018A&A...618L..10G}, where three hot spots were detected at different epochs. These structures stay stable in terms of the luminosity for one orbital period or a large fraction of it. The small flux density changes can be attributed solely to the Doppler boosting along an orbit, which has a large inclination of $\sim 160^{\circ}$ but is not face-on (see also Fig.~\ref{fig_illustration_MAD} for the inclination scheme). Flares that were found to be closer to edge-on orbits often exhibit a peak-shoulder structure in their light curves, which can be attributed to the combination of the gravitational lensing and the Doppler boosting along their orbits \citep{2017FoPh...47..553E,2017MNRAS.472.4422K}. There is also no evidence of significant shearing along this orbit during one orbital period.

Keeping this in mind, we assume the following properties for the hot spot.  
\begin{itemize}
\item[(a)] The hot spot is a bound test, potentially charged mass moving on a circular orbit in the Kerr spacetime background, as well as in the global poloidal magnetic field.
\item[(b)] Given the lack of any shearing along the orbit, we also assume that the hot spot is a spherical mass, or at least that its shape does not change during one orbital timescale.
\item[(c)] We also assume that during the orbital timescale, the surrounding environment does not affect the hot spot dynamics, or, in other words, these effects are negligible in comparison with the general relativistic effects and the electromagnetic interaction.
\end {itemize}
Assumption (a) will be justified in more detail in Sections~\ref{subsec:MF}, \ref{sec:BHcharge}, \ref{sizeandmass}, and \ref{subsec_mag_field_hotspot}. Circular trajectories should be preferred, since the collisions of magnetically confined cloudlets in the MAD models on crossing orbits are highly dissipative in the central 100 $R_{\rm g}$ \citep[see also][for the comparison of circular and elliptical hot spot trajectories]{1994ApJ...425...63B,2018arXiv180600284E}. 

Concerning assumption (b), there is no observational information about the shape of the hot spot. However, given the stable intrinsic luminosity and no evidence for shearing, a stable spherical shape is a reasonable assumption. The shape of the hot spot may also be kept stable against magnetohydrodynamic instabilities by the tangled internal magnetic field \citep{2015MNRAS.449....2M,2017ApJ...834L..19G}, which is also the property of MADs. 

As a new feature in comparison with previous studies of the hot spot dynamics and radiative properties \citep{2005MNRAS.363..353B,2006ApJ...636L.109B,2006MNRAS.367..905B,2006A&A...460...15M,2008JPhCS.131a2008Z,2010A&A...510A...3Z}, we add the potential electromagnetic interaction. Any other effects from the surrounding medium or objects (stars) may have a certain effect, but it is observationally difficult to estimate them at the moment as the hot spots are not observed beyond one orbital period. We leave the other potential magnetohydrodynamic effects for future studies when more data are available.

In addition, the accretion flow close to Sgr A* is diluted and potentially clumpy, as predicted by MADs \citep[][]{2003PASJ...55L..69N,2008ApJ...677..317I}, which we illustrate in Fig.~\ref{fig_illustration_MAD} and discuss in more detail in Section~\ref{subsec: flare-hot spot} and \ref{sec:discussion}. Therefore, it is likely that the hot spot is a dominating mass on a circular orbit close to the ISCO. This is also supported by the observations of discrete X-ray and NIR flares that point toward clumpy accretion. Given the general instability of the accretion flow below the magnetospheric radius (see Equation~(\ref{eq_magnetospheric_radius})), the flow continues inward in the form of magnetically confined blobs that diffuse through the poloidal magnetic field via reconnection events. In this way, a hot spot is a dominating mass at a given time and location close to the ISCO; hence, any effect from the surrounding medium is assumed to be negligible during one orbital timescale. The surrounding medium has the nature of a hot, diluted corona that is heated up by the released energy of the MAD flow \citep{2003PASJ...55L..69N}. In fact, the number density of the blob is expected to be larger than the mean accretion flow density by the inverse of the volume filling factor -- $(r_{\rm ISCO}/R_{\rm hs})^3<216$ -- which we derive in Section~\ref{subsec: flare-hot spot}.

In addition, even if the surrounding medium with a comparable number density were present around the hot spot, it would be comoving with the hot spot close to the ISCO. The magnetohydrodynamic drag force can be expressed as \citep{2008ApJ...677..993D,2015MNRAS.449....2M}
\begin{equation}
    F_{\rm drag}\sim \rho_{\rm a} v_{\rm rel}^2 R_{\rm hs}^2 \left(1+\frac{v_{\rm A}^2}{v_{\rm rel}^2}\right)\,,
    \label{eq_drag_force}
\end{equation}
where $\rho_{\rm a}$ is the ambient density, $v_{\rm rel}$ is the relative velocity between the hot spot and the ambient medium, $R_{\rm hs}$ is the hot spot radius, and $v_{\rm A}$ is the Alfven velocity. The drag force is therefore negligible for the case where the ambient medium is comoving with the hot spot, since $v_{\rm rel}\approx 0$. This justifies the assumption (c).

It is likely that shearing and the associated departure from the quasi-spherical shape govern the hot spot evolution after one orbital period when the hot spot flux density falls beyond the detection limit. However, at the same time, the hot spot likely plunges from the ISCO toward and beyond the event horizon of Sgr A* on a timescale of $t_{\rm rf}=12GM/c^3\approx 4\,{\rm minutes}$ for the radial fall towards Sgr~A* (in general, this depends on the initial angular momentum). The shearing and infall would manifest themselves by the gradually decreasing flux densities of the flare \citep[see the simulated light curves of an infalling, shearing hot spot calculated by][using KYSPOT code]{2004ApJS..153..205D,2017Obs...137..267Z}, which has not been detected so far \citep[see the observed light curves in Figs. 1 and 2 in][]{2018A&A...618L..10G}. This so-far-unobserved regime is therefore beyond the scope of the current paper but may be of interest in our future studies.

In other words, the dynamical effects studied in this work focus on the transient hot spot feature during its orbital timescale. Since this timescale, which can be calculated for the nonrotating black hole at the ISCO as 
\begin{equation}
 P_{\rm hs}=30.3 \left(\frac{M}{4\times 10^6\,M_{\odot}}\right) \left(\frac{r}{6R_{\rm g}} \right)^{3/2}\,{\rm minutes},
\end{equation}
is much shorter than the viscous timescale of the hot and thick accretion flow around Sgr~A* at larger scales of $100\,R_{\rm g}$,
\begin{equation}
    t_{\rm visc}\sim 2.3 \left(\frac{M}{4\times 10^6\,M_{\odot}}\right)\left(\frac{h}{r}\right)^{-2}\left(\frac{r}{100 R_{\rm g}}\right)^{3/2} \left(\frac{\alpha_{\rm visc}}{0.1}\right)^{-1}\,{\rm days}\,,
\end{equation}
where we assumed a thick flow with a height-to-radius ratio of $h/r\sim 1$ and   a viscosity parameter of $\alpha_{\rm visc} \sim 0.1$. Therefore, we can neglect the long-term behavior of the whole flow as such. These long-term effects will be studied in more detail in our upcoming studies.

%Here we need a derivation of the hot spot mass and size, and general discussion of other parametetrs, like number density, temperature and so on.
\subsection{Magnetic Field Estimates} \label{subsec:MF}

Current estimates of the magnetic field around Sgr~A* -- at the scales of the ISCO -- are consistent with its strength of the order of $10-100$G \citep{2012A&A...537A..52E}. The time-variable flux density during high states -- flares -- is modelled using synchrotron components in the NIR domain \citep{2012ApJS..203...18W,2018ApJ...863...15W}. To explain the emission mechanism of simultaneous X-ray and NIR flares \citep[X-ray flares always have NIR counterparts, but not vice versa;][]{2016A&A...589A.116M}, synchrotron -- sychtrotron-self-Compton (SYN-SSC) model is often employed that requires relativistic electrons with a Lorentz factor of $\gamma_{\rm e}\sim 10^3$ \citep{2012A&A...537A..52E}, which is 3 orders of magnitude less than the pure synchrotron-synchrotron model. The time-lag of $t=1.5 \pm 0.5\,{\rm h}$ between X-ray/NIR and millimeter/submillimeter flares \citep{2008A&A...492..337E} is successfully explained by comoving adiabatic expansion of plasma blobs with uniform expansion speeds of $v_{\rm exp}\sim 0.005-0.017\,{\rm c}$  \citep[][{
 see also the model description in Subsection~\ref{subsec: flare-hot spot}}]{2010A&A...517A..46K,2008ApJ...682..361Y}. The SSC model can also be used to estimate the magnetic field strength using $B\sim \theta_{\rm ss}^4 \nu_{\rm m}^5 S_{\rm m}^{-2}$, where $\theta_{\rm ss}$ is the angular source size and $S_{\rm m}$ is the flux density at the turnover frequency $\nu_{\rm m}$. Typical values of the magnetic field during high states as derived from the SSC modeling are of the order of $B\sim 10-100\,{\rm G}$ \citep{2012A&A...537A..52E,2010A&A...517A..46K}, in general being variable by a factor of a few.

Another constraint can be derived from the Faraday rotation measurements and the accretion flow density and temperature profiles close to Sgr~A* that are inferred from fitting the RIAF model to the X-ray observations of the hot flow. Based on the Faraday rotation measurements of the magnetar PSR J1745-2900, \citet{2013Natur.501..391E} put a lower limit of $B\gtrsim 8\,{\rm mG}$ on the line-of-sight component of the magnetic field for the magnetar deprojected distance of $r\gtrsim 0.12\,{\rm pc}$ from Sgr~A*. They also confirmed that the ordered magnetic field is present at length-scales of $\sim 0.1\,{\rm pc}$, which is intermediate between the large-scale ordered magnetic field in the central molecular zone \citep{2015llg..book..391M} and the ordered field close to the ISCO of Sgr~A* \citep{2015Sci...350.1242J,2018A&A...618L..10G}. The plasma magnetization parameter can be expressed as the ratio of its magnetic field energy density to its thermal pressure (thermal pressure of electrons), $\beta_{\rm p}=B^2/8\pi n_{\rm p} k_{\rm B}T_{\rm p}$, where $n_{\rm p}$ and $T_{\rm p}$ are the plasma density and temperature, respectively. { The value of $\beta_{\rm p}$ at the distance of the magnetar ($r \sim 0.1\,{\rm pc}$) can be estimated based on the density and the temperature as inferred for the Bondi radius, $n_{\rm p}\approx 26\,{\rm cm^{-3}}$ and $T_{\rm p}\approx 1.5\times 10^{7}\,{\rm K}$ \citep{2003ApJ...591..891B}. The value of the magnetic field is inferred from the Faraday rotation, $B\sim 8\,{\rm mG}$ \citep{2013Natur.501..391E}. Then the magnetization parameter is $\beta_{\rm p}\sim 47.3$. To estimate $\beta_{\rm p}$ at ISCO, we adopt the values of the density, the temperature, and the magnetic field from the plasmon model applied to the radio variability by \citet{2006ApJ...650..189Y}, who got $n_{\rm p}\sim 6 \times 10^5\,{\rm cm^{-3}}$, $T_{\rm p}\sim 10^{9}\,{\rm K}$, and $B\sim 10\,{\rm G}$. Then the magnetization parameter at the ISCO can be estimated as $\beta_{\rm p}\sim 48.1$. Between the Bondi radius and the ISCO, the magnetization parameter may thus be considered constant within the uncertainties. Its value of $\beta_{\rm p}\sim 50$ also implies that the plasma in this region is magnetically dominated.}

The X-ray spectroscopy measurements by the Chandra telescope \citep{2013Sci...341..981W} revealed an elongated extended emission structure with a radius of $1.5''$ centered at Sgr~A*. Based on the very weak Fe K$\alpha$ line, a no-outflow scenario can be rejected. The radial density profile $n_{\rm p}\propto r^{-3/2+s}$ with $s\gtrsim 0.6$ best fits the continuum and the emission lines in the $2-10$ keV band. Using the upper limit on the mass accretion rate by Sgr~A*, $\dot{M}_{\rm SgrA*}\sim 2\times 10^{-7}\,M_{\odot}{\rm yr^{-1}}$ \citep{2007ApJ...654L..57M}, and the assumption of the inner radius of RIAF at $r_{\rm i}\sim 200\,R_{\rm g}$, \citet{2013Sci...341..981W} got a radial temperature profile $T_{\rm p}\propto r^{-\theta_{T}}$ with $\theta_{T}\gtrsim 0.6$. Using the thermal and magnetic pressure coupling via the magnetization parameter, $P_{\rm mag}=\beta_{\rm p}P_{\rm th}$, { and under the assumption of the constant $\beta_{\rm p}$}, we obtain the power-law scaling of the magnetic field strength, $B\propto r^{-3/4+1/2(s-\theta_{T})}$, which for $s \approx \theta_{T}$ { \citep[as inferred from X-ray spectra by][]{2013Sci...341..981W}} simply becomes $B\propto r^{-3/4}$. We can normalize the magnetic field profile using the line-of-sight magnetic field, as well as the magnetar distance from \citep{2013Natur.501..391E} \\

\begin{equation}
    B(r) \gtrsim 8\times 10^{-3} \left(\frac{r}{5.2\times 10^{5}\,R_{\rm g}} \right)^{-3/4}\,{\rm G}\,,
    \label{eq_mag_field}
\end{equation}
where the distance was scaled to $r\simeq 0.1\,{\rm pc} \sim 5.2 \times 10^5\,R_{\rm g}$. The radiating plasma components are approximately at a distance of $r=10\,R_{\rm g}$, which implies a magnetic field of $B(10\,R_{\rm g})\gtrsim 28\,{\rm G}$. This is consistent with the magnetic field as inferred from flare observations \citep{2012A&A...537A..52E}. The Bondi flow (no outflow) with $s=0$ and $\theta_{T}\sim 1$ would lead to a radial dependency $B\propto r^{-5/4}$ and $B(10\,R_{\rm g})\gtrsim 6300\,{\rm G}$, which is 2 orders of magnitude larger than the flare value. This gives further support to the general RIAF model with an outflow, where $\lesssim 1\%$ of the material captured at the Bondi radius is accreted by Sgr~A*. The presence of an outflow then leads to the flattening of the density profile. \\

\subsection{Limits on the black hole charge} \label{sec:BHcharge}

In \citet{2018MNRAS.480.4408Z}, we used the current observations of the hot phase of Sgr A* surroundings within the innermost arcsecond to place constraints on the electric charge of Sgr~A*. The existence of a hot quasi-neutral, stationary plasma around Sgr~A* leads to the existence of the equilibrium charge $Q_{\rm eq}$, which stops the separation of lighter electrons from heavier protons in collisionless plasma. The equilibrium charge may be expressed as
\begin{equation}
    Q_{\rm  eq}^{\eta_{T}}=\frac{4\pi \epsilon_0 G}{e}\left(\frac{\eta_{T} m_{\rm p}-m_{\rm e}}{1+\eta_{T}}\right)M\,,
    \label{eq_equilibrium charge}
\end{equation}
where $\eta_{T}\equiv T_{\rm e}/T_{\rm p}\simeq 1-1/5$, i.e. the hot flow close to the black hole is characterized by different proton and electron temperatures, with the proton temperature up to five times larger than the electron temperature \citep{2009ApJ...706..497M,2010ApJ...717.1092D}. For $\eta_{T}=1$, we obtain $Q_{\rm eq}^1=3.1\times 10^8\,C$ and for $\eta_{T}=1/5$, the charge is lower by about a factor of three, $Q_{\rm eq}^{1/5}=1.02\times 10^8\,C$. In some models, much smaller values of $\eta_{T}$ are adopted. For example, \citet{2016A&A...586A..38M} considered the values formally down to $\eta_{T}^{\rm min}=1/100$ for a highly magnetized accretion flow, which would lead to an equilibrium charge of $Q_{\rm eq}^{1/100}=5.85\times 10^6$C. On the other hand, at least in the accretion flow part, there are
strong arguments for a considerable part of the heating going directly to electrons \citep[see, e.g.,][]{1997ApJ...486L..43B,2018A&A...615A..57M}, which prevents such a small value of $T_{\rm e}/T_{\rm p}$. 

In a more general case, the black hole rotation in an ordered, homogeneous magnetic field $B_{\rm ext}$ leads to the twisting of the magnetic field lines and the generation of an electric field associated with an induced Wald charge, $Q_{\rm W}=2a M B_{\rm ext}$ \citep{1974PhRvD..10.1680W}, with respect to the infinity, where $a$ is a  dimensionless spin parameter.  The dimensionless spin parameter $a$ is defined using the relation $a_{\rm spin}=aGM/c^2$, with $a=1$ standing for the maximum prograde rotation, $a=-1$ representing a maximally counterrotating black hole, and $a=0$ being a nonrotating black hole. Considering the constraint on the spin, $a \leq M$, the upper limit on the induced Wald charge for the Galactic center black holes is 
\begin{equation}
    Q_{\rm W}\leq 2.3\times 10^{15}\left(\frac{M}{4 \times 10^6\,M_{\odot}}\right)^2\left(\frac{B_{\rm ext}}{10\,{\rm G}}\right)\,{\rm C}\,.
    \label{eq_charge_wald}
\end{equation}
The expected black hole charge based on the realistic magnetohydrodynamic environment is in the range $Q=(10^8, 10^{15})\,{\rm C}$, which is at least 12 orders of magnitude below the extremal value,
\begin{equation}
    Q_{\rm max}=6.9 \times 10^{26}\left(\frac{M}{4\times 10^6\,M_{\odot}}\right)\sqrt{1-a^2}\,{\rm C}\,.
\end{equation}

One should stress that the Galactic center black hole is not in a vacuum which is assumed by the Wald solution. A force-free approximation may be more appropriate. However, \citet{PhysRevD.98.123002} showed that a black hole embedded in the force-free magnetosphere is expected to carry charge; hence, a black hole carrying a small charge seems more likely than a completely neutral black hole. 

A charged black hole that rotates will itself generate a dipole magnetic field with a magnetic dipole moment $m_{\rm d}\sim Q_{\rm W}M$. The dipole magnetic field strength is then given by $B_{\rm d}\sim m_{\rm d}/r^3=2B_{\rm ext} a M^2/r^3$, with the upper limit given by the maximally rotating black hole, $a < M $ and $r_{\rm ISCO}=M$, which gives $B_{\rm d}^{\rm max}(r=r_{\rm ISCO})<2B_{\rm ext}$.

\subsection{Size and mass of the flare components} \label{sizeandmass}

{The parameters} of the flare components, such as size and mass, can be { estimated} using the results of current and previous flare studies from Sgr~A* observed in millimeter, NIR, and to X-ray parts of spectra \citep{2012A&A...537A..52E}. Observed flares have strengths of $5.2$mJy in NIR K-band that is $40\%$ of S2 star having corresponding strengths of $13$mJy. A simple estimate of the length scale of the hot spot may be derived from the adiabatic expansion of the emitting sources observed at speeds of $\sim 0.01c$ \citep{1974ApJ...192..261J}. Light-travel arguments give constraints on size of the flare components as 
\begin{equation} \label{Rhs1}
    R_{\rm hs} \approx  R_{\rm g} \frac{ G M}{c^2} \approx 6\times 10^{11}    \left(\frac{M}{4 \times 10^6 M_\odot} \right) {\rm cm}. 
\end{equation}
Number densities of flare components can be estimated to be of  the order of 
\begin{equation} \label{rhoN1}
\rho_{\rm N} \approx 10^{7\pm1} {\rm cm}^{-3}  ,  
\end{equation}
derived from both the synchrotron model and radio Faraday rotation of the polarization planes \citep{2012A&A...537A..52E,2003ApJ...598..301Y}.  For pure electron and electron-proton cases with spherical hot spot, we get the following limits on the masses of the flare components:
\begin{equation} \label{masshs1}  
m_{\rm hs}^{\rm min} \approx  8.7 \times 10^{14} {\rm g}; \quad
m_{\rm hs}^{\rm max} \approx 1.6 \times 10^{20} {\rm g}\,.   
\end{equation}
For the comparison, the solar mass is $M_{\odot} \approx 2 \times 10^{33}$g, and the typical masses of large asteroids are of the order of $10^{23}$g. 
 
 Another estimate comes from the SSC modelling of the simultaneous NIR and X-ray flares. The source properties of these flares may be constrained using the power-law energy distribution with an exponential cutoff, $N(\gamma)=N_0\gamma^{-p}\exp{(-\gamma/\gamma_{\rm c})}$ (with $p\sim 2$). In this model, NIR flares are produced via the synchrotron mechanism with electrons gyrating in  the magnetic field, and the same electrons upscatter the emitted photons to higher energies, producing the X-ray emission. With the knowledge of the magnetic field strength and assuming that the hot spot is uniform and spherical, one obtains the radius and number density of the same order of magnitude \citep{2007gsbh.book.....M},
 
 \begin{align}
     R_{\rm hs} & \approx 5.12\left(\frac{L_{\rm syn}}{10^{36}\,{\rm erg\,s^{-1}}}\right)\left(\frac{L_{\rm SSC}}{10^{35}\,{\rm erg\,s^{-1}}}\right)^{-1/2}\times \nonumber \\ & \times  \left(\frac{B_{\rm ext}}{10\,{\rm G}}\right)^{-1}\,R_{\rm g}\,, \label{Rhs2}
     %\nonumber 
     \\
     \rho_{\rm N} & \approx  1.15 \times 10^6\left(\frac{L_{\rm syn}}{10^{36}\,{\rm erg\,s^{-1}}}\right)^{-2}\left(\frac{L_{\rm SSC}}{10^{35}\,{\rm erg\,s^{-1}}}\right)^{3/2}\times \nonumber \\ & \times  \left(\frac{B_{\rm ext}}{10\,{\rm G}}\right)\left(\frac{\gamma_{\rm c}}{100} \right)^{-2}\,{\rm cm^{-3}}\,,%\nonumber
     \label{rhoN2}
 \end{align}
 which, by the order of magnitude,  is close to the estimates given by Equations (\ref{Rhs1}) and (\ref{rhoN1}).  Here we define by $L_{\rm syn}$ and $L_{\rm SSC}$ the luminosities of the hot spot in the synchrotron and synchrotron-self-Compton regimes, respectively. 
 This leads to the hot spot mass of $m_{\rm hs}^{\rm min}\sim 1.2\times 10^{17}\,{\rm g}$ and $m_{\rm hs}^{\rm max}\sim 2.3\times 10^{20}\,{\rm g}$ for the pure electron and proton limits, respectively.

\subsection{Magnetic Field Influence on the Motion of the Hot Spot}
\label{subsec_mag_field_hotspot}

In order to test the influence of electromagnetic interaction on the dynamics of the hot spot around Sgr~A*, one can consider a simplified scenario of the motion of a charged test particle around a rotating black hole in the presence of a magnetic field. It is natural (and supported by the observed orthogonality of the magnetic field lines and the hot spot orbital plane) to assume that the magnetic field shares the symmetries of the spacetime in the vicinity of {a} black hole. Therefore, using the stationarity and axial symmetry of the Kerr black hole spacetime, one can express the four-vector potential in terms of the time-like and space-like Killing vectors $\xi^\mu$ in the form
\begin{equation}
    A^{\mu} = C_1 \xi_{(t)}^{\mu} + C_2 \xi_{(\phi)}^{\mu}, \label{VecPot}
\end{equation}
where $C_1$ and $C_2$ are constants. The solution (\ref{VecPot}), corresponding to the test field approximation was suggested by \cite{1974PhRvD..10.1680W}, and in case of asymptotically uniform magnetic field with the strength $B$ the nonvanishing components of $A_\mu$ correspond to
\begin{equation} \label{fourvecuni}
    A_t = \frac{B}{2} \left(g_{t\phi} + 2 a g_{tt}\right), \quad A_{\phi} =  \frac{B}{2} \left(g_{\phi\phi} + 2 a g_{t\phi}\right).
\end{equation}
This is also the historically first analytical solution of Maxwell equations in the Kerr spacetime background. If the field has an inclination angle with respect to the spin axis of a black hole, the solution for $A_\mu$ is given by \cite{1985MNRAS.212..899B}. A configuration of a magnetic field corresponding to the dipole type has been derived by  \cite{Petterson:1974:PHYSR4:}. In all axially symmetric electromagnetic field configurations, the energy and angular momentum of test particle with charge $q$ and mass $m$ is modified according to \citep{2016PhRvD..93h4012T}
\begin{eqnarray}
    - {\cal E} \equiv - \frac{E}{m} = % \xi^{\mu}_{(t)} \frac{P_{\mu}}{m} = 
    g_{tt} \frac{dt}{d\tau} + g_{t\phi}\frac{d\phi}{d\tau} + \frac{q}{m} A_{t},&& \label{Energy} \\
{\cal L} \equiv \frac{L}{m} = % \xi^{\mu}_{(\phi)} \frac{P_{\mu}}{m} = 
g_{\phi\phi} \frac{d\phi}{d\tau} + g_{t\phi}\frac{dt}{d\tau} + \frac{q}{m} A_{\phi}.&& \label{AngMom} 
\end{eqnarray}
 The most general form of the equations of motion for charged particles in curved spacetime is given by the DeWitt--Brehme  equation \citep{DeW-Bre:1960:AnnPhys:,Hobbs:1968:AnnPhys:}, which can be written in the form
\begin{eqnarray} 
&& \frac{D u^\mu}{d \tau} = \frac{q}{m} F^{\mu}_{\,\,\,\nu} u^{\nu} 
+ \frac{2 q^2}{3 m} \left( \frac{D^2 u^\mu}{d\tau} + u^\mu u_\nu \frac{D^2 u^\nu}{d\tau} \right) \nonumber \\ 
&& + \frac{q^2}{3 m} \left(R^{\mu}_{\,\,\,\lambda} u^{\lambda} + R^{\nu}_{\,\,\,\lambda} u_{\nu} u^{\lambda} u^{\mu} \right) + \frac{q^2}{m} ~f^{\mu \nu}_{\rm \, tail} \,\, u_\nu, 
\label{eqmoDWBH}  
\end{eqnarray}
where $R^{\mu}_{\nu}$ is  the Ricci tensor, $F_{\mu\nu} = A_{\nu,\mu} - A_{\mu,\nu}$ is the Faraday tensor, $D$ denotes a covariant derivative, and four-velocity $u^{\mu}=dx^{\mu}/d\tau$ satisfies the condition $u^\mu u_\mu = -1$. The last term in Eq. (\ref{eqmoDWBH}), known as the tail integral, reads 
\begin{equation}
f^{\mu \nu}_{\rm \, tail}  = \int_{-\infty}^{\tau}     
D^{[\mu} G^{\nu]}_{ + \lambda'} \bigl(\tau,\tau'\bigr)   
u^{\lambda'}(\tau') \, d\tau' ,
\end{equation}
where $G^\mu_{+\lambda}$ is a retarded Green's function. 
A detailed analysis of this equation, together with numerical integration in astrophysically relevant situation, can be found in \cite{2018ApJ...861....2T,2018AN....339..341T}. In the realistic conditions, the leading forces acting on a charged test particle are the Lorentz force and the radiation reaction force, given by the first and second terms on the right-hand side of Eq.(\ref{eqmoDWBH}). The terms containing Ricci tensors are irrelevant if the Kerr spacetime metric is assumed, while the tail term is negligible in comparison to the rest of the terms. Using the approach by \cite{Lan-Lif:1975:CTF:}, we get the covariant dynamical equations of the motion of charged test particle in curved spacetime in the presence of an electromagnetic field,
\begin{equation} \label{eq-LL1}
\frac{D u^\alpha}{d \tau} = \frac{q}{m} F^{\alpha}_{\,\,\,\beta} u^{\beta} + \frac{2 q^3}{3 m^2} f^\alpha_R,
\end{equation}
where 
\begin{equation}   \label{eq-LL2}
f^\alpha_R =  \frac{D F^{\alpha}_{\,\,\,\beta}}{d x^{\mu}} u^\beta u^\mu + \frac{q}{m} \left( F^{\alpha}_{\,\,\,\beta} 
F^{\beta}_{\,\,\,\mu} +  F_{\mu\nu} F^{\nu}_{\,\,\,\sigma} u^\sigma u^\alpha \right) u^\mu .
\end{equation}
Equation (\ref{eq-LL1}) with equation (\ref{eq-LL2}) is the covariant form of the Landau-Lifshitz equation describing the dynamics of radiating charged particle. We will use this equation for constraints on the motion of hot spots.

\section{Limits to the charge of flare components} \label{sec:charge-limit}

\subsection{Charge separation in a plasma surrounding Sgr~A*} \label{sec:GJ-argument}

It is usually assumed that a { plasma surrounding astrophysical black holes} is electrically neutral due to neutralization of charged plasma { on} relatively short timescales. Any oscillation of {the net} charge density in a plasma is supposed to disappear very quickly due to induction of a large electric field caused by charge imbalance. However, in the presence of an external magnetic field and when the plasma is moving at relativistic speeds, one can observe the charge separation effect in a magnetized plasma and consequently measure the net charge density. Applied to rotating neutron stars with magnetic fields, this special relativistic effect of plasma charging is known as the Goldreich-Julian (GJ) charge density \citep{1969ApJ...157..869G}. In fact, the motion of a plasma induces an electric field that, in the comoving frame of a plasma, should be neutralized, which leads to the appearance of the net charge in the rest frame. The GJ charge density is usually referred  to  for pulsar magnetospheres, although it is applicable in more general cases as well, as we will show below. For a black hole magnetosphere, the charging of the plasma was first described by \cite{1975PhRvD..12.2959R}, who showed that the twisting of magnetic field lines due to rotation of the black hole induces an electric charge in both the black hole and surrounding magnetosphere with equal and opposite signs of the charge value. 

\subsubsection{Special Relativistic Case}

Neglecting for now the general relativistic effects, Maxwell's equations read
\begin{eqnarray}
 \nabla \cdot {\bf E} = 4\pi \rho, &\quad& 
 \nabla \cdot {\bf B} = 0,\label{eq-Max_a} \\
 \nabla \times {\bf E} = -\frac{\partial {\bf B}}{\partial t }, &\quad& 
 \nabla \times {\bf B} = 4 \pi {\bf j} + \frac{\partial {\bf E}}{\partial t }.\label{eq-Max_b}
\end{eqnarray}
where $\rho$ and ${\bf j}$ are the charge and current densities. 
For a frame moving with a system with the velocity $v$ with respect to the rest frame, the Lorentz transformations lead to
\begin{eqnarray}
 {\bf E'} &=& \gamma ({\bf E} + {\bf v} \times {\bf B}) - \frac{\gamma^2}{\gamma + 1} {\bf v} ({\bf v} \cdot {\bf E}), \label{eq-Max1}
 \\ 
 {\bf B'}   &=& \gamma ({\bf B} + {\bf v} \times {\bf E}) - \frac{\gamma^2}{\gamma + 1} {\bf v} ({\bf v} \cdot {\bf B}), \label{eq-Max2}
 \end{eqnarray}
where "primes" denote the quantities measured with respect to the inertial frame moving at the given moment together with the system and   $\gamma = (1-v^2)^{-1/2}$. In a comoving frame of the system, the current density is connected with the electric field by Ohm's law, ${\bf j'} = \sigma {\bf E'}$, where $\sigma$ is the conductivity of the medium. For an observer at rest, one gets the Ohm's law in the form
\begin{equation}
 {\bf j} = \gamma \sigma  \left( {\bf E} + {\bf v} \times {\bf B} - {\bf v} \cdot 
 ({\bf v} \, %\cdot 
 {\bf E}) \right) + \rho {\bf v}.
\end{equation}
Let us now assume that the matter containing plasma is a perfect electrical conductor. This implies that the following relation holds:
\begin{equation} \label{eq-evB}
    {\bf E} = - {\bf v} \times {\bf B},
\end{equation}
From this, it follows that an external observer measures the induced electric field that arises in order to compensate the electric field in the comoving frame of the system. 

One can express  the velocity in terms of an orbital angular velocity of the hot spot moving around the black hole in the equatorial plane, ${\bf v} = {\bf \Omega} \times {\bf R}$. Substituting Eq. (\ref{eq-evB}) into the first equation of Eq. (\ref{eq-Max_a}) in terms of angular velocity and dividing to elementary charge $e$, we get the net charge number density in a plasma (number density of extra electrons or protons) in the form
\begin{equation} \label{rhoqGJ}
    \rho_q = \frac{1}{2 \pi c}  \frac{\Omega \, B_{\perp}}{ |e|} ,
\end{equation} 
where $\Omega$ is the orbital angular velocity of the hot spot and $B_{\perp}$ is the strength of the magnetic field orthogonal to the orbital plane.  The orbital period of the hot spots at the distance of the innermost stable circular orbit (ISCO $\sim 6 G M/c^2$) of Sgr~A* is $T \sim 45$minutes, which corresponds to the angular velocity $\Omega = 2\pi/T \sim 2.33 \times 10^{-3} s^{-1}$. The equipartition strength of the magnetic field at the ISCO scale can be assumed to be of the order of $10$G \citep{2017FoPh...47..553E}. 
Thus, the number density of extra charged charged particles is 
\begin{equation} \label{rho-gj-sr}
    \rho_q \approx 2.57 \times 10^{-4} \left(\frac{B}{10{\rm G}} \right) \left( \frac{T}{45 \rm{min}}\right)^{-1} {\rm cm^{-3}},
\end{equation}
which is at least $10^{10}$ times less than the total number density in a plasma, given by Eq.(\ref{rhoN1}). 
Assuming a spherical volume of the radius $R \sim R_{\rm g}$, corresponding to the size estimate (\ref{Rhs1})  we get a net charge excess of the order of
\begin{equation} \label{qGJ1}
    |q| \approx  4 \times 10^{13} \left(\frac{B}{10{\rm G}} \right) %\left(\frac{v}{v_{\rm isco}} \right) 
    \left(\frac{R}{R_s}\right)^3 
    {\rm C}
\end{equation}
One can see that the charge separation in a relativistic  magnetized plasma can lead to the presence of a sufficient net charge in the hot spots that can considerably affect their motion. The dynamics of a charged hot spot is discussed in Section~\ref{sec:sub32}.

Let us now find the limits to the ratio of the Lorentz and gravitational forces acting on the hot spot at the ISCO scales, assuming that the hot spot has a charge given by Eq.(\ref{qGJ1}) and the magnetic field is orthogonal to the orbital plane with a strength of $10$G. For a hot spot with a mass in the  range given by Eq.(\ref{masshs1}), moving with a velocity $v\sim 0.3 c$ around Sgr~A* (as observed in recent GRAVITY flares) at a distance of $\sim 6R_{\rm g}$, we get the following limits: 
\begin{equation} \label{florfgrav}
   10^{-5} < \frac{F_{\rm Lor.}}{F_{\rm grav.}} < 10.  
\end{equation}
We will give tighter constraints on this ratio  in Section~\ref{sec:sub32} by analyzing the period--radius relations of the hot spot orbits (parameterizing the above ratio by the dimensionless parameter ${\cal B}$) and comparing them with those of the three most recent flares.

\subsubsection{General Relativistic Case}

It is important to note that the general relativistic version of Eq.(\ref{rhoqGJ}) leads to { a} similar order estimate as Eqs.(\ref{rho-gj-sr}) and (\ref{qGJ1}) unless a hot spot is moving in the very close vicinity of the event horizon. However, for completeness, we derive the net charge density of the flare component due to charge separation in a magnetized plasma in curved spacetime following the works of  \cite{1972ApJ...178..347B,1975PhRvD..12.2959R,1982MNRAS.198..339T,1992MNRAS.255...61M,2004MNRAS.350..407I}. 
For our purposes, the most convenient way to describe the electrodynamics of relativistic plasma around a black hole is to use the approach of 3 + 1 splitting of spacetime introduced by \cite{1982MNRAS.198..339T} and further  developed in \cite{2004MNRAS.350..407I}. 
In curved spacetime, the Maxwell equations read in a similar way as in flat-space case (Eqs. (\ref{eq-Max_a}) and (\ref{eq-Max_b})), except that the operator $\nabla$ is taken in 3D curved coordinates, implying a covariant derivative of absolute space. In tensor form, the covariant Maxwell equations read
\begin{equation} 
  \nabla_\nu \,  ^*\!F^{\mu \nu} = 0, \quad \nabla_\nu  F^{\mu \nu} =  J^\mu, 
\label{Maxw1}
\end{equation}
where $F^{\alpha\beta}$ and $^*\!F^{\alpha\beta}$ are the Maxwell and Faraday tensors, respectively, and $J^\mu$ is the four-current.  
Splitting these equations into time and space components, we get 
\begin{eqnarray}
 \nabla \cdot {\bf B} = 0,
   &\quad& 
\nabla \times {\bf E} = -\frac{\partial {\bf B}}{\partial t }, \label{eq-MaxGR_a} \\
 \nabla \cdot {\bf D} = 4\pi \rho, &\quad& 
 \nabla \times {\bf H} = 4 \pi {\bf j} + \frac{\partial {\bf D}}{\partial t }.\label{eq-MaxGR_b}
\end{eqnarray}
It should be noted that ${\bf D}$ and ${\bf H}$ coincide with ${\bf E}$ and ${\bf B}$ measured by a zero angular momentum observer (ZAMO), whose four-velocity in axially symmetric spacetime is defined by
\begin{equation}
     n^{\mu} = %
     (n^t,0,0,n^\phi), 
\end{equation}
where
\begin{equation}
 (n^t)^2=\frac{g_{\phi\phi}}{g_{t\phi}^2 %
  -g_{tt}g_{\phi\phi}}, %
  \quad 
  n^\phi=%
  -\frac{g_{t\phi}}{g_{\phi\phi}}\,n^t.
\end{equation}
Applying the covariant derivative $\nabla$ to Eq.(\ref{eq-MaxGR_b}) we get the charge conservation law
\begin{equation}
    \partial_{t}\rho + \nabla \cdot {\bf{J}}= 0.
\end{equation}
Assuming that the magnetosphere of a black hole shares the background symmetry of the black hole, i.e. applying stationarity and  axial symmetry, we get the effective charge density in the form %\cite{1992MNRAS.255...61M}
\begin{eqnarray}
    \rho = - \frac{1}{4 \pi} \nabla \cdot \left[ \frac{1}{\alpha} \left(1- \frac{k }{\eta_{r}^3} \right) {\bf v} \times {\bf B}   \right], \\
    {\bf v} = {\bf \Omega} \times {\bf r}, \quad k = \frac{R_{\rm g} \beta}{ a }, \quad \eta_{r} = \frac{a}{r}, \quad \alpha = \sqrt{g_{tt}},
\end{eqnarray}
where $a$ is the black hole's  spin parameter, $\beta$ is the moment of inertia of the plasma rotating around the black hole, and $\alpha$ is the lapse function.

It was argued by \cite{Komissarov:2004ms} that an electric field measured by ZAMO drives the electric current along the magnetic field lines, resulting in the separation of charges and the drop % 
of the electrostatic potential, at least within the ergosphere. In this scenario, { the} black hole can act as the unipolar generator \citep{1977MNRAS.179..433B} similar to the classical Faraday disk, which is based on the use of electromotive force $q {\bf v} \times {\bf B}$, resulting in the charge separation due to the voltage drop between the edge of the disk and its center. Further analysis led to conclusion that any rotating compact object, like neutron stars or black holes, immersed into an external magnetic field and surrounded by plasma or an accretion disk   generates a rotationally induced electric field at the object, as well as in the surrounding magnetosphere. 

\begin{figure*}
  %\centering
  \includegraphics[width=0.49\textwidth]{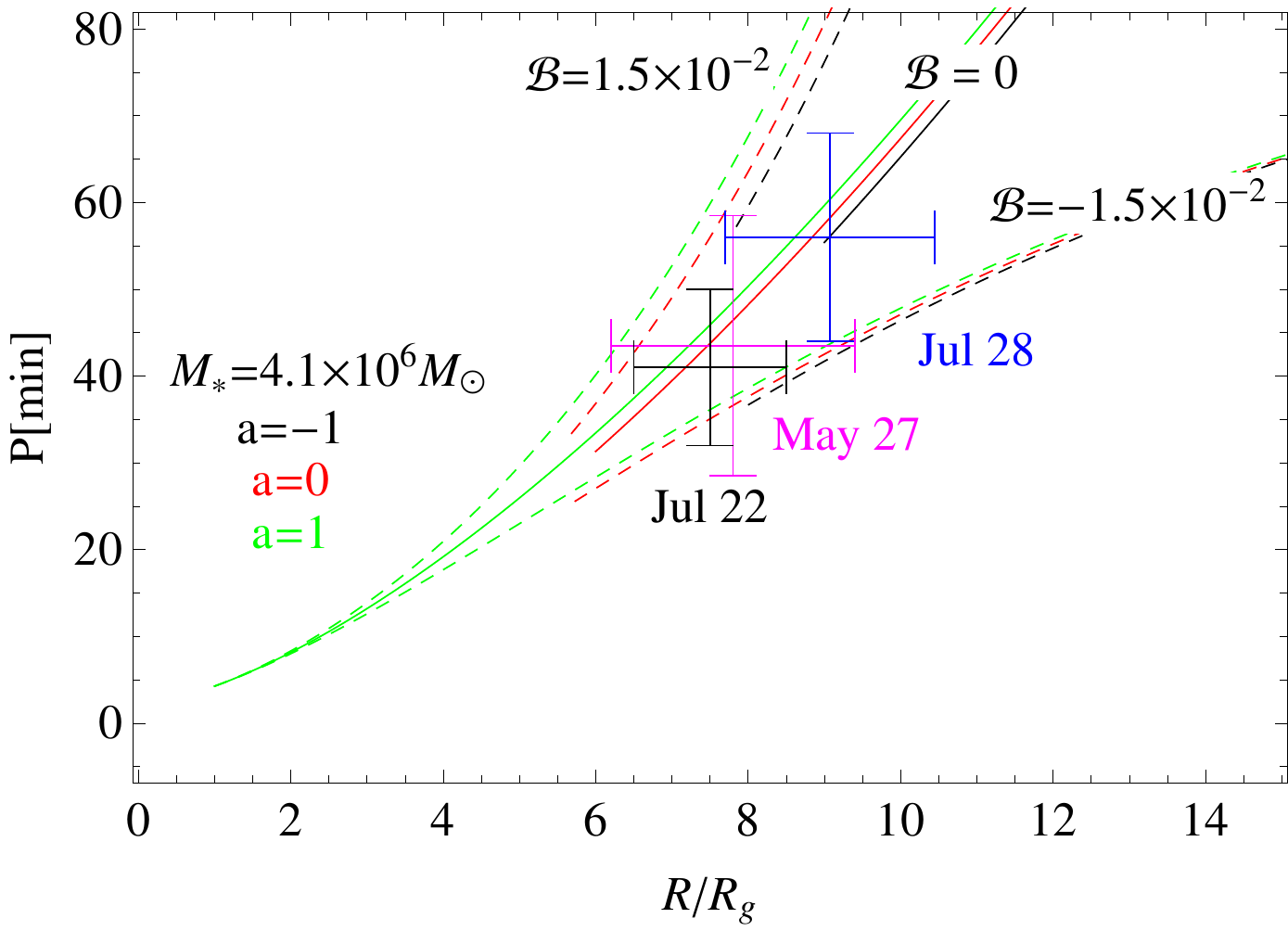}
  \includegraphics[width=0.49\textwidth]{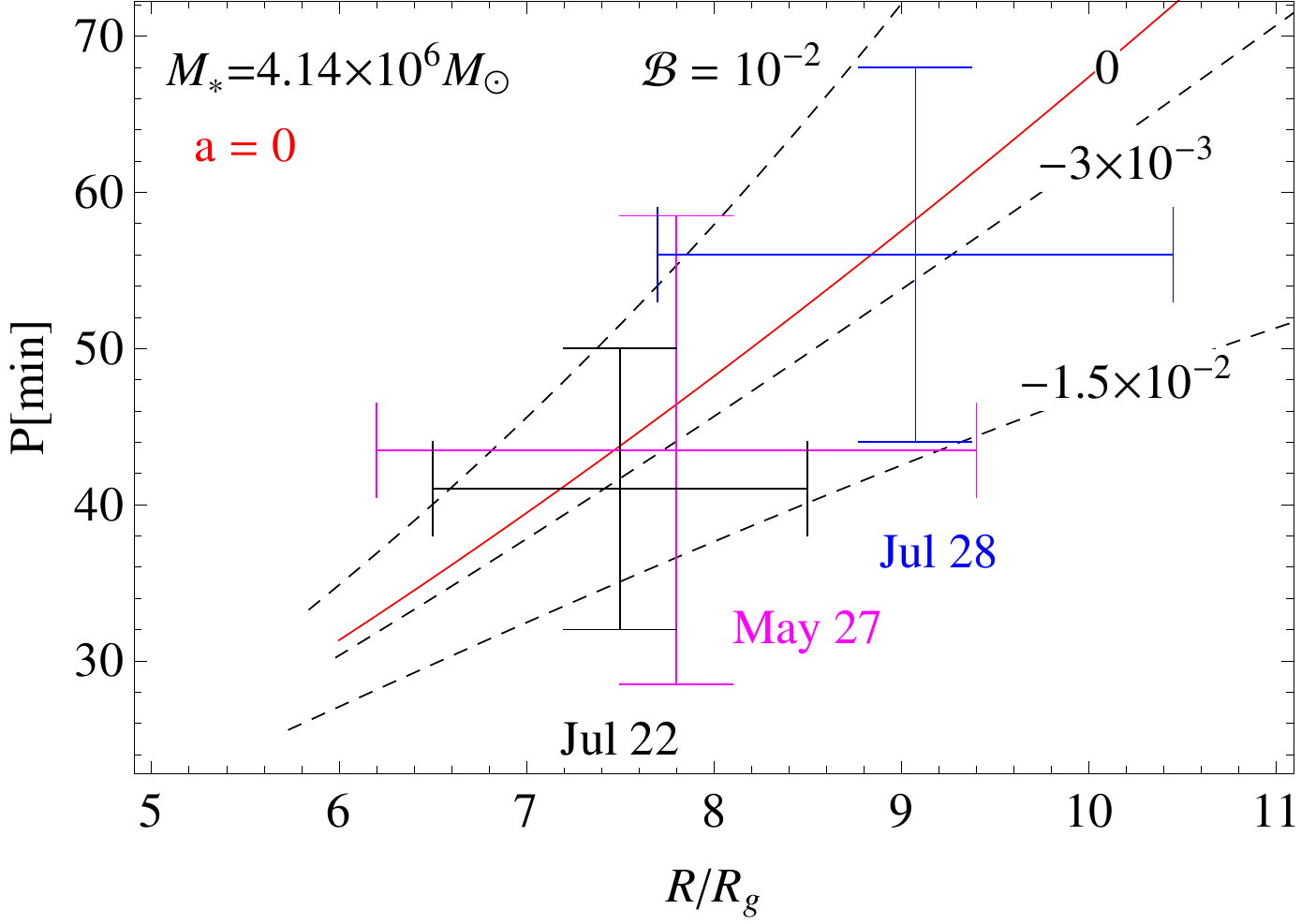}
  \caption{Left: Orbital period--radius relations of three flares observed by GRAVITY on July 22 (black), May 27 (pink), and July 28 (blue) fitted with circular orbits of a charged hot spot moving around a Kerr black hole of mass $4\times10^6M_\odot$ immersed in an external magnetic field orthogonal to the orbital plane and characterized by the parameter ${\cal B}$, given by Eq.(\ref{BBparam}).  Solid lines correspond to the ${\cal B}=0$ case, describing orbits without electromagnetic interaction, while dashed lines correspond to the limiting values of ${\cal B}=\pm 1.5\times 10^{-2}$, fitting the observed periods and positions of the flares. Green and black curves  correspond to the extremal Kerr black hole with $a=\pm 1$, and red curves correspond to the Schwarzschild black hole with $a=0$. Right: same as the left plot but zoomed in for the nonrotating black hole ($a=0$). The centers of the error bars for all three flares can be fitted by the parameter ${\cal B}= - 3\times 10^{-3}$ (middle dashed line).
}
  \label{fig1}
\end{figure*}

\subsection{Dynamics of the charged hot spots} \label{sec:sub32}

  Let us start with a note on terminology. There have been a number of papers over the last three decades developing a phenomenological description of an electromagnetic signal from a hot spot orbiting near a black hole \citep[see, e.g.][and references cited therein]{2004ApJ...606.1098S,2006AN....327..961K}. This notion has been rather successful in predicting basic features that appear due to general relativity in light curves, spectra, and polarimetrical signal that are expected from a spatially localized source on a circular orbit or a plunging trajectory near the event horizon. Various physical representations have been invoked in order to understand the extended life span of the spot in the presence of strong tidal forces, where a plain blob of gas would disintegrate on a timescale shorter than the orbital period (which would prevent clear signatures from showing up and set the constraints of black hole parameters); the term "spot" can thus represent a stable vortex \citep{1992Natur.356...41A} or a wave pattern \citep{2001PASJ...53..189K} within the accretion medium, it can be stabilized by the presence of a stellar object in the core \citep{1973ApJ...183..237C,1994ApJ...425...63B,2014A&A...565A..17Z,2015ApJ...800..125V}, or it can be confined by ambient magnetic pressure.

 Although black holes do not support their own magnetic fields, large-scale organized fields are possible and even likely to occur due to external currents flowing in the accretion medium. Magnetic effects can visibly the influence the motion and radiation of an electrically charged spot. Even if a plasma blob is electrically neutral globally, the mechanism of charge separation produces an excess of charge that can prevail in certain regions.

The motion of a charged hot spot is described by Eq.(\ref{eq-LL1}), which also includes in addition to the geodesic term, the Lorentz and radiation reaction forces. In the case of a locally uniform magnetic field that satisfies the solution given by (\ref{fourvecuni}), the relative influence of magnetic and gravitational fields on the motion of the hot spot can be parameterized by the dimensionless parameter \citep{2018ApJ...861....2T,2016PhRvD..93h4012T,2010PhRvD..82h4034F}
\begin{equation} \label{BBparam}
    {\cal B} = \frac{q_{\rm hs}\,G\,B\,M}{ 2\, m_{\rm hs} \, c^4 },
\end{equation}
where $B$ is the strength of the magnetic field, $M$ is the black hole mass, $q_{\rm hs}$ and $m_{\rm hs}$ are the charge and the mass of the hot spot, and $G$ and $c$ are constants. The factor $1/2$ is given for historical reasons. Hereafter, we call $\BB$ the magnetic parameter. The parameter $\BB$ is chosen in such a way that $2 \BB v / c$ is the ratio of the Lorentz force to the gravitational force acting on the hot spot and constrained for a plasma surrounding Sgr~A* in Eq.(\ref{florfgrav}). 

Following \cite{2018A&A...618L..10G}, we assume that the magnetic field is  orthogonal to the orbital plane of the hot spots that is placed at the equatorial plane of spinning Sgr~A*. Depending on the orientation of the Lorentz force, the shift of the orbital frequency can occur in both { directions} for {a} given value of the orbital radii. 
Given the period--radius relations of components of three flares observed in 2018 July 22, May 27, and July 28, we solve the equations of motion for the charged hot spot numerically and put constraints on the magnetic parameter $\BB$ in Figure~\ref{fig1} as $-0.015< {\cal B}< 0.01$. One can also see in Figure~\ref{fig1} (right) that the positions of the centers of the observed flares on the period--radius plot are slightly lower than the theoretically predicted periods of neutral hot spots (red curve). The value of the magnetic parameter ${\cal B}$ fitting the mean values (centers) of observed periods and radii of three flares is ${\cal B} \sim -3 \times 10^{-3}$ (dashed middle curve).   
As discussed in Section~\ref{subsec:MF}, various measurements and estimates of magnetic field strength in the vicinity of Sgr~A* suggest the equipartition magnetic field with a strength of $B\sim10$G \citep{2017FoPh...47..553E}. This gives limits on the specific charge (charge-to-mass ratio) following Figure~\ref{fig1}, in the range 
\begin{equation} \label{qtilde1}
\frac{|q_{\rm hs}|}{m_{\rm hs} } <  10^{-3} \left(\frac{B}{10 {\rm G}}\right)^{-1} 
 \left(\frac{M_{\rm SgrA^*}}{4\times 10^{6} M_{\odot}}  \right)^{-1}  {\rm C/g}, 
\end{equation} 
while the mean value of the specific charge fitting the observed mean periods and radii corresponds to
\begin{equation} \label{qtilde2}
    \frac{q_{\rm hs}}{m_{\rm hs} } \bigg|_{\rm mean} \approx - 3 \times 10^{-4} \left(\frac{B}{10 {\rm G}}\right)^{-1} 
 \left(\frac{M_{\rm SgrA^*}}{4\times 10^{6} M_{\odot}}  \right)^{-1}  {\rm C/g}.
\end{equation}
The same ratio for electrons is of order $e/m_{e}\sim - 10^8$C/g and for protons  $e/m_{p} \sim 10^5$C/g. Assuming that the constituents of the flare components are mainly protons and electrons, one can easily calculate the limiting ratio of the number of extra net charged particles to the number of the neutral particles (proton--electron pairs) in the hot spot as
\begin{eqnarray}
    \frac{N_{\rm charged}}{N_{\rm neutral}} = \frac{q_{\rm hs}}{e} \,\, \frac{m_{\rm p} + m_{\rm e}}{m_{\rm hs}}   < 10^{-8}, \\ 
    \frac{N_{\rm charged}}{N_{\rm neutral}} \bigg|_{\rm mean} \approx 3 \times 10^{-9},
\end{eqnarray}
i.e., observed flare components have a net charge concentration corresponding to 1 extra charged particle to at least $10^8$ neutral pairs of protons and electrons.  Since the mean value of the specific charge given by Eq.(\ref{qtilde2}) is negative, this corresponds to an extra electron in each $3\times 10^8$ neutral pair. Based on the total number density obtained in Eqs.~(\ref{rhoN1}) and (\ref{rhoN2}), one can estimate the number density of extra charged particles ($\rho_q  = \rho_N \, {N_{\rm charged}}/{N_{\rm neutral}}$) as 
\begin{eqnarray}
    \rho_{\rm q} < 10^{-2} \left(\frac{B}{10 {\rm G}}\right)^{-1} 
 \left( \frac{\rho_{\rm N}}{ 10^{6} \, {\rm cm^{-3}}}  \right)   {\rm cm^{-3}}, \label{rho-dyn}\\
    \rho_{\rm q}^{\rm mean} \approx 3 \times 10^{-3} 
     \left(\frac{B}{10 {\rm G}}\right)^{-1} 
 \left( \frac{\rho_{\rm N}}{ 10^{6} \, {\rm cm^{-3}}}  \right)   {\rm cm^{-3}}, \label{rho-mean}
\end{eqnarray}
that is, an order of magnitude larger than our earlier estimate (Eq.(\ref{rho-gj-sr})), which is based on the charge separation in a relativistic magnetized plasma. However, the discrepancy can be easily omitted if one assumes a slightly stronger magnetic field at the orbital location of the hot spots of the order or less than $\lesssim 100$G. In that case, the two approaches will perfectly match. 

For the masses of the hot spots estimated in Eq.(\ref{masshs1}), the limiting values for the charges of the hot spots are
\begin{eqnarray} \label{qhsdyn1}
 |q_{\rm hs}| < 10^{14.5 \pm 2.5} \left(\frac{B}{10 {\rm G}}\right)^{-1 } 
 \left(\frac{m_{\rm hs}}{10^{17.5 \pm 2.5} {\rm g}}\right) {\rm C} \,, %\nonumber 
 \\
 q_{\rm hs}^{\rm mean} \approx  - 10^{13.5 \pm 2.5} \left(\frac{B}{10 {\rm G}}\right)^{-1 } 
 \left(\frac{m_{\rm hs}}{10^{17.5 \pm 2.5} {\rm g}}\right) {\rm C} \,,  \label{qhsmean1}
 %\nonumber 
\end{eqnarray}
which does not contradict the value of Eq.(\ref{qGJ1}) estimated above. 
It is important to note that according to Figure~\ref{fig1}, the spin parameter of the black hole does not play a crucial role in the fitting of the period--radius relation of circular orbits, while the magnetic parameter shifts the orbits and periods significantly, although  one can conclude that the plasma containing the hot spot has a nonnegligible excess of net charge, which is shown by the above estimates. 

\subsection{Synchrotron Radiation from Charged Hot Spot} \label{sec:synchrotron}

Emission of flares observed mainly at X-ray wavelengths roughly appears every day, increasing the luminosity of Sgr~A* up to 2 orders of magnitude \citep{2003ApJ...598..301Y}. On average, the flare state of Sgr~A* corresponds to a luminosity of the order of $10^{33}$erg s$^{-1}$, although the brightest flares may reach a luminosity of $10^{35}-10^{36}$erg s$^{-1}$ \citep{2012ApJ...759...95N}. The flare activity states may last from a few minutes to hours. Usually, the shortest timescales correspond to the flare components at  the closest distances from Sgr~A*. Associating the nonthermal flare emission with the synchrotron radiation of a charged hot spot in a magnetic field surrounding Sgr~A*, one can find other limits on the charge of the flare components and their emission timescales. For a magnetic field orthogonal to the orbital plane, the intensity of radiation in all directions of the hot spot orbiting the black hole in a fully relativistic approach is given by \citep{Sok-Gal-Pet:1978:PLA:,Shoom:2015:PHYSR4:,2018ApJ...861....2T}
 \begin{equation} \label{rad-par}
     L = \frac{2}{3} \frac{q_{\rm hs}^4 B^2 v^2 \gamma^2}{m_{\rm hs}^2 c^3} \left(1-\frac{2 R_{\rm g}}{R_0}\right)^3 {\rm erg \,\,s^{-1}},
 \end{equation}
where $v$ is the velocity of the hot spot in units of the speed of light, and $R_0$ is the orbital radius. Equalizing Eq.(\ref{rad-par}) to $10^{33}$erg s$^{-1}$ for the orbit at the radius $R=6R_{\rm g}$ from Sgr~A*, we get another limit for the charge of the hot spot as $q_{\rm hs}^{\rm min} < q < q_{\rm hs}^{\rm max}$, where
\begin{equation} \label{qhs-synch1}
 q_{\rm hs}^{\rm min} \approx - 10^{13} \left(\frac{B}{10 {\rm G}}\right)^{-\frac{1}{2} } 
 \left(\frac{m_{\rm hs}}{10^{14} {\rm g}}\right)^{\frac{1}{2} }
 \left(\frac{L}{10^{33} {\rm erg \, s^{-1}}}\right)^{\frac{1}{4} } {\rm C}, 
\end{equation}
\begin{equation} 
q_{\rm hs}^{\rm max} \approx  10^{16} \left(\frac{B}{10 {\rm G}}\right)^{-\frac{1}{2} } 
 \left(\frac{m_{\rm hs}}{10^{20} {\rm g}}\right)^{\frac{1}{2} }
 \left(\frac{L}{10^{33} {\rm erg \, s^{-1}}}\right)^{\frac{1}{4} } {\rm C}, \label{qhs-synch2}
\end{equation}
whose orders of magnitude are very close to the limits of Eq.(\ref{qhsdyn1}) given by the fitting of the period--radius data.  

\subsection{Effect of the Black Hole Charge} \label{sec:bhcharge-dynamics}

\begin{figure}
  %\centering
  \includegraphics[width=1\hsize]{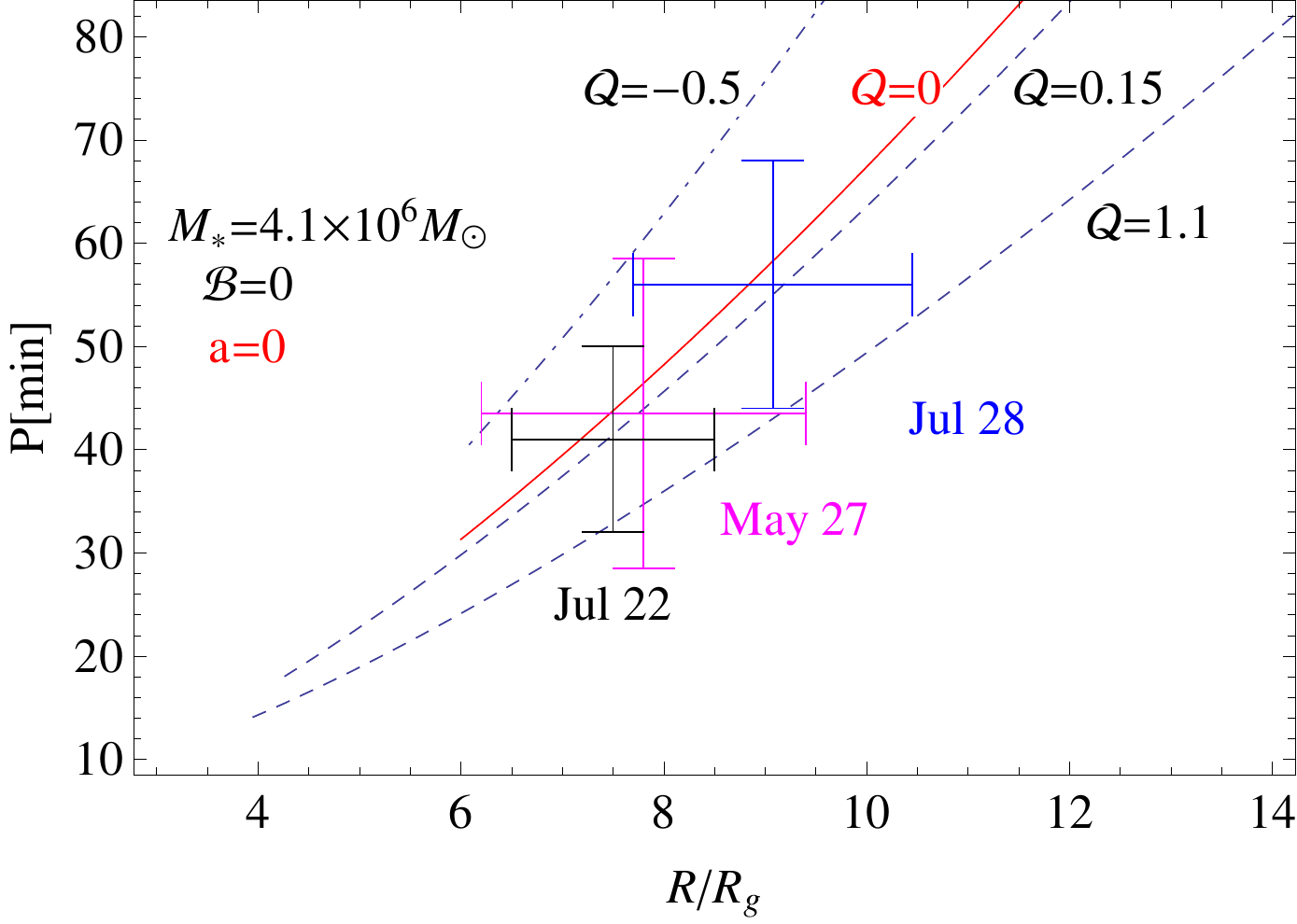}
  \caption{ { Orbital period--radius relations of three flares observed by GRAVITY on July 22 (black), May 27 (pink), and July 28 (blue) fitted with circular orbits of a charged hot spot moving around a Schwarzschild black hole of mass $4\times10^6M_\odot$ carrying a small electric charge. The dynamics of the hot spots is characterized by the parameter ${\cal Q}$ defined by Eq. (\ref{param-Q}) and reflecting the Coulombic interaction between the hot spot and the black hole. The centers of the error bars for all three flares can be fitted by the mean value of parameter ${\cal Q}= 0.15$ (middle dashed line). }
}
  \label{fig2}
\end{figure}

Given that the hot spot carries a small electric charge, one can also consider the possible influence on its motion of the unscreened charge of the black hole, which is discussed in Section~\ref{sec:BHcharge}. The realistic upper limit to the black hole charge is of the order of  $\sim 10^{15}$C \citep[see][]{2018MNRAS.480.4408Z}, which can arise due to magnetic field twist. In the case of a uniform magnetic field, the black hole possesses a Wald charge $Q_{\rm W} = 2 G^2 a M B/c^4 \approx G^2 M^2 B/ (2 c^4)$.
Using an approach similar to the magnetic field case (see Section~\ref{sec:sub32} for details) we summarize the results of the fitting of the period--radius plots in Figure~\ref{fig2}. For the sake of simplicity and in order to identify the pure contribution to the hot spot dynamics due to black hole's charge, we neglect the effects of the spin and magnetic field. Similar to the magnetic parameter given by Eq.(\ref{BBparam}), we introduce the dimensionless parameter ${\cal Q}$, reflecting the Coulombic interaction between the hot spot and the black hole, 
\begin{equation} \label{param-Q}
    {\cal Q} = \frac{q_{\rm hs} \, Q_{\rm BH}}{G\, m_{\rm hs}\, M_{\rm SgrA^*}}.
\end{equation}
Taking the charge of the black hole $Q_{\rm BH} \sim 10^{15}$C and constraining the charge parameter to  $-0.5 < {\cal Q} < 1$, with a mean value ${\cal Q} =0.15$, we get the following constraints on the specific charge (charge-to-mass ratio) of the hot spot: 
\begin{equation} \label{qtilde3}
\frac{|q_{\rm hs}|}{m_{\rm hs} } <  2 \times 10^{-2} \left(\frac{Q}{10^{15} {\rm C}}\right)^{-1} 
 \left(\frac{M_{\rm SgrA^*}}{4\times 10^{6} M_{\odot}}  \right)^{-1}  {\rm C/g},
\end{equation} 
while the mean value of the specific charge fitting the observed mean periods and radii corresponds to
\begin{equation} \label{qtilde4}
    \frac{q_{\rm hs}}{m_{\rm hs} } \bigg|_{\rm mean} \approx 7 \times 10^{-3} \left(\frac{B}{10 {\rm G}}\right)^{-1} 
 \left(\frac{M_{\rm SgrA^*}}{4\times 10^{6} M_{\odot}}  \right)^{-1}  {\rm C/g}.
\end{equation}
One can notice that the estimates in Eqs.(\ref{qtilde3}) and (\ref{qtilde4}) are only 1 order of magnitude larger than in the magnetic case  (see Eqs. (\ref{qtilde1}) and (\ref{qtilde2})). Further analysis leads to the following constraints to the charge: 
\begin{eqnarray} \label{qhs-QQ1}
 q_{\rm hs}^{\rm min} \approx - 10^{12} \left(\frac{Q}{Q_{\rm W} (\sim 10^{18}C)}\right) 
 \left(\frac{m_{\rm hs}}{10^{17} {\rm g}}\right)
 {\rm C}, \\
 q_{\rm hs}^{\rm max} \approx  10^{18} \left(\frac{Q}{Q_{\rm W} (\sim 10^{15}\rm C)}\right)
  \left(\frac{m_{\rm hs}}{10^{20} {\rm g}}\right)
 {\rm C}, \label{qhs-QQ2}
\end{eqnarray}
which are close to the previous estimates by an order of magnitude.

\subsection{ISCO Shifts Mimicking Black Hole Spin} \label{sec:isco-shift}

\begin{figure*}
  %\centering
  \includegraphics[width=\hsize]{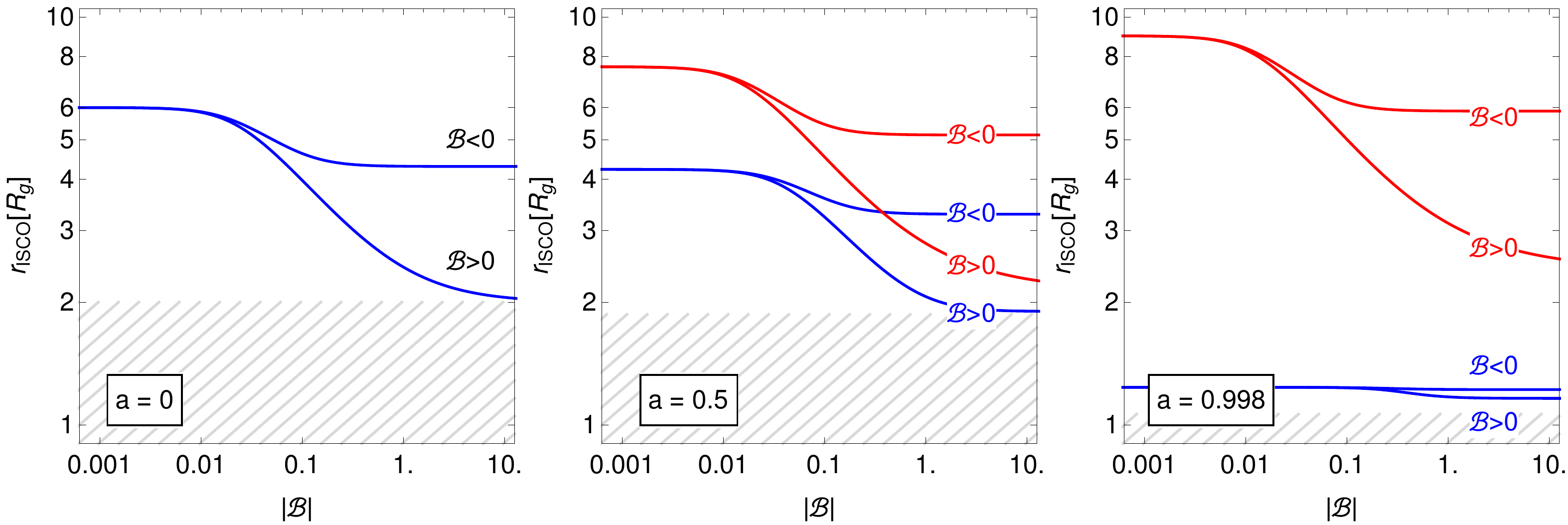}
  \caption{Location of the ISCO of a positively and negatively charged hot spot corotating (blue) and counterrotating (red) around Sgr~A* immersed in a  uniform magnetic field in dependence on the dimensionless parameter ${\cal B} = q G M_{\rm BH} B/(m_{\rm hs} c^4)$  for the values of black hole spin: $a=0$, $a=0.5$, $a=0.998$.   The charge of the black hole corresponds to the Wald charge.  The region below the event horizon is shown by gray hatched lines. 
}
  \label{figisco}
\end{figure*}

The location of the innermost stable circular orbit (ISCO) is among the few parameters that are strongly sensitive to the value of the black hole spin. The ISCO of a nonrotating black hole ($a=0$) is located at a distance of $6R_{\rm g}$ from singularity. For rotating black holes, the ISCO of corotating matter shifts toward the black hole, coinciding with the event horizon at   the extremal case ($a=1$). For counterrotating matter, the ISCO shifts outward from the black hole, reaching up to $9R_{\rm g}$ in the extremal case.  However, the inclusion of the interaction of the magnetic field with the charge of the accretion flow can shift the ISCO dramatically. The motion of relativistic plasma around magnetized black hole puts limits to the ratio of the Lorentz force to the gravitation force acting at the ISCO scales, given by Eq. (\ref{florfgrav}). This implies that in general, for a plasma, the upper limit for the magnetic parameter ${\cal B}$ defined in Eq.~(\ref{BBparam}) is $|{\cal B}|<10$. 

In Figure~\ref{figisco} we demonstrate the shift of the ISCO location toward the black hole by increasing the magnetic parameter ${\cal B}$ in the case of uniform magnetic field configuration. The ISCO in the presence of a magnetic field has in total four branches by two for each corotating and counterrotating cases corresponding to the Larmor and anti-Larmor types of  motion \citep{2002MNRAS.336..241A,2010PhRvD..82h4034F,2016PhRvD..93h4012T}. For the upper limit of the magnetic parameter $|{\cal B}|\approx 10$, the ISCO in the nonrotating black hole case shifts to the values $r_{\rm ISCO} \approx 2.1 R_{\rm g}$ for ${\cal B}>0$  and $r_{\rm ISCO} \approx 4.3 R_{\rm g}$ for ${\cal B}<0$. These values correspond to the ISCO of neutral matter moving around rotating black hole with the spin $a=0.93$ and $a=0.48$, respectively. 

A simple example demonstrated in Figure~\ref{figisco} shows that the effect of electromagnetic interaction on the matter surrounding the black hole can be of crucial importance, as it may lead to the discrepancy in the measurements of the spin of the SMBH. Since the magnetic field configuration in the Sgr~A* environment can be more sophisticated, the problem requires further study. 

Similar analyses of the location of the ISCO in the case of the Schwarzschild black hole with a small electric charge immersed in the external magnetic field were studied in \cite{2019arXiv191107645H}.

\section{Inclination of the Black Hole Spin} \label{sec:sub35}

\begin{figure*}
  %\centering
  \includegraphics[width=\hsize]{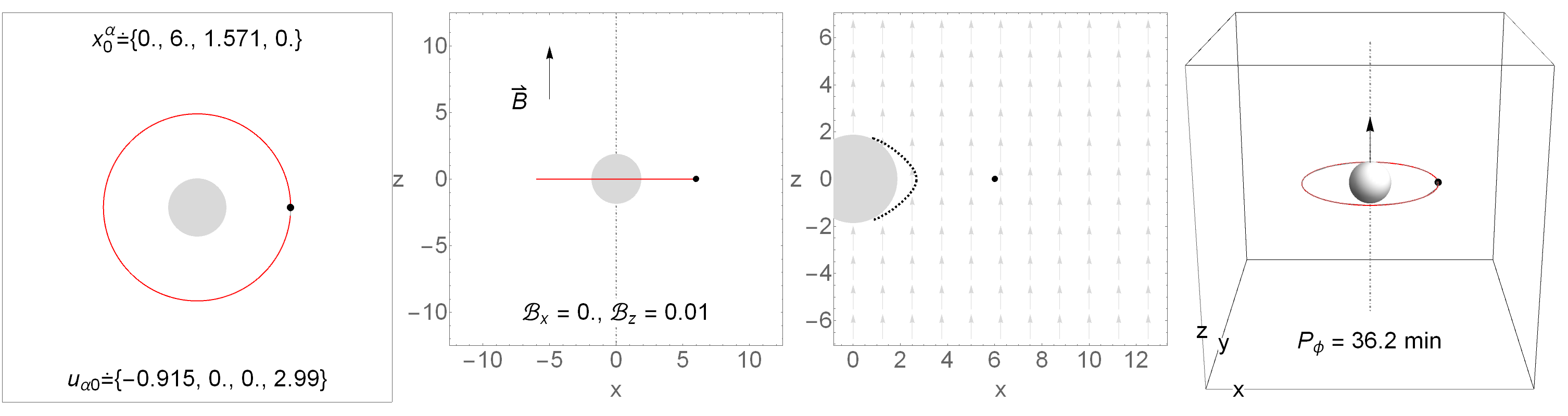}
  \includegraphics[width=\hsize]{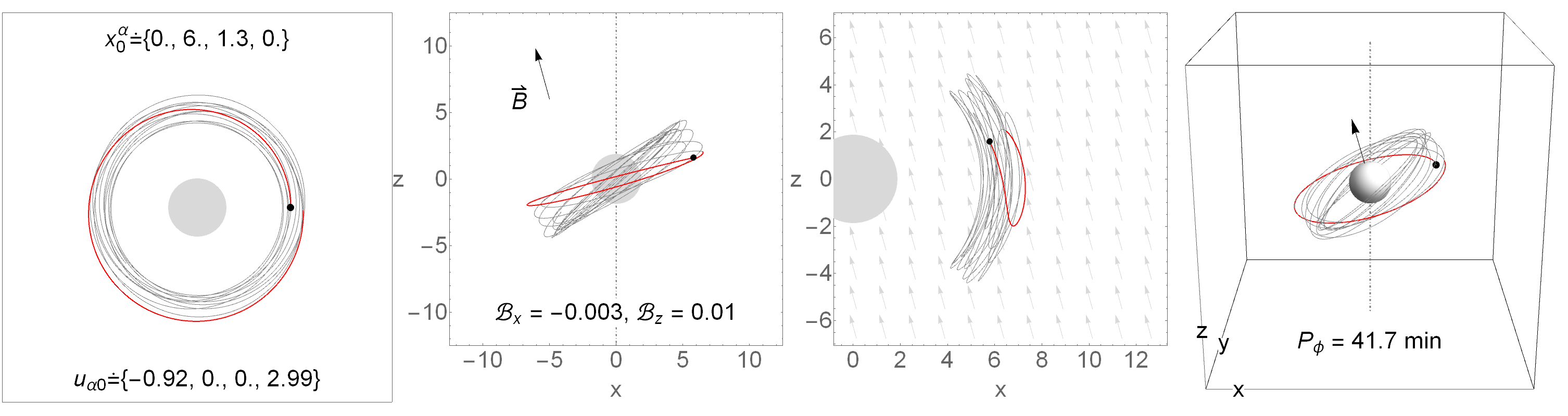}
  \includegraphics[width=\hsize]{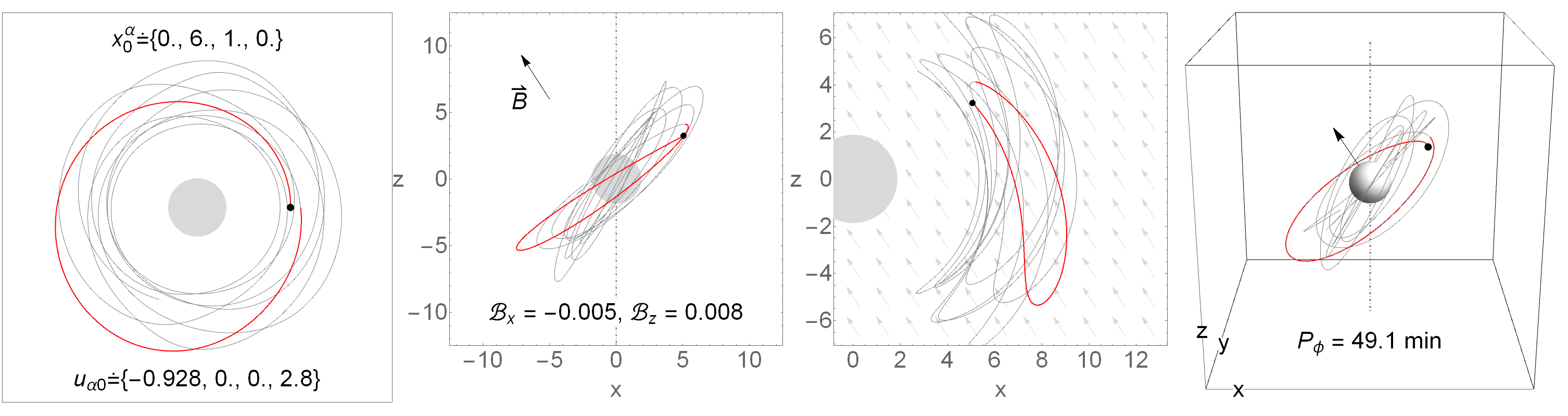}
  \includegraphics[width=\hsize]{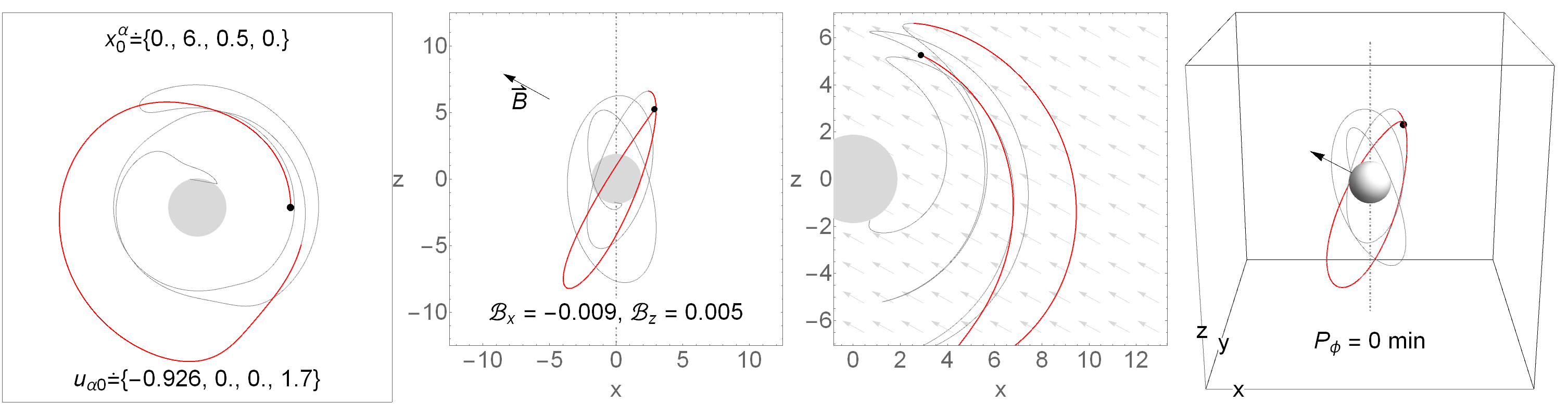}
  \caption{Trajectories of charged hot spots orbiting SMBH Sgr~A* for various relative inclinations of magnetic field and the black hole spin axes. The orbital plane is orthogonal to the magnetic field lines, the spin is fixed to the value $a=0.5M$ and vertically directed, the initial positions and velocities are given inside the plots and are the same for all plots. The first column represents a face-on view of the orbit; the second column shows the trajectories viewed from the equatorial plane, with the dot--dashed line indicating the axis of black hole spin. The third column shows the cross section of the trajectory in a plane orthogonal to the equatorial plane, and the fourth column represents the 3D trajectories of the hot spots with the orbital periods found numerically.  See the detailed discussion in Section \ref{sec:sub35}.
}
  \label{fig3}
\end{figure*}

\begin{figure*}
  %\centering
  \includegraphics[width=\hsize]{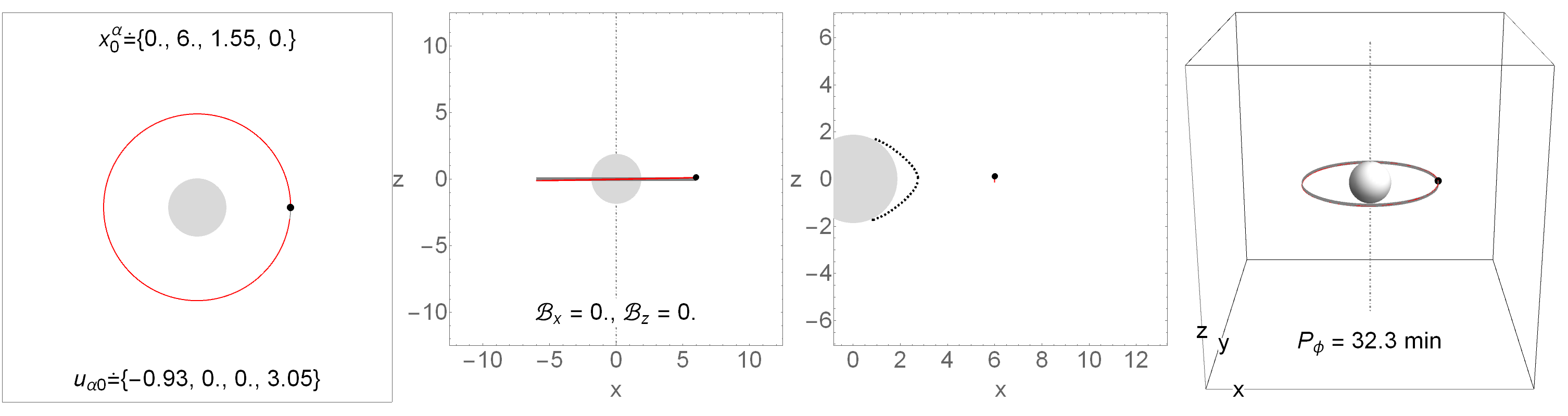}
  \includegraphics[width=\hsize]{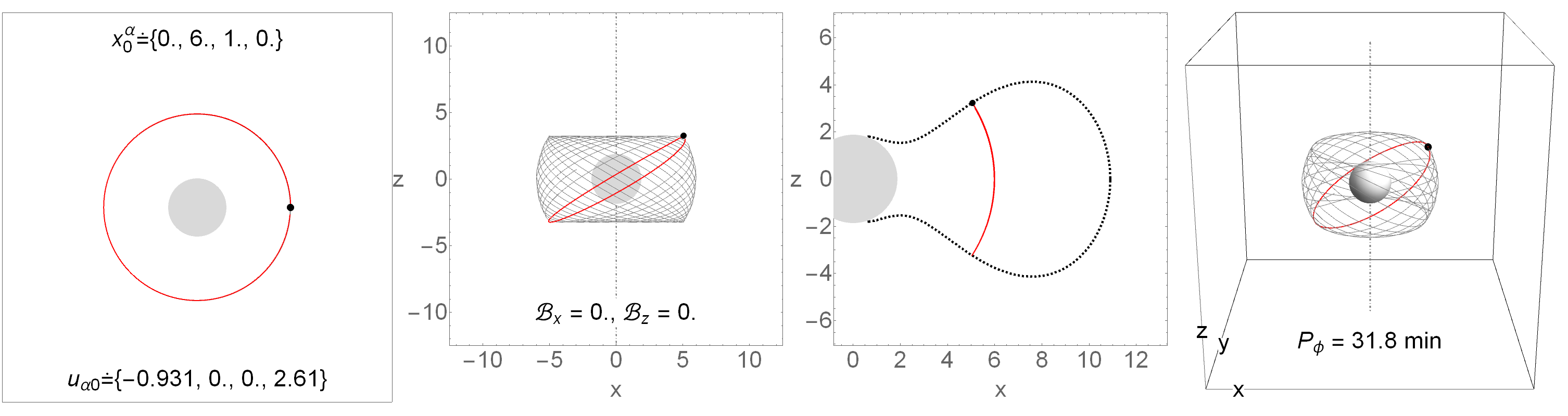}
  \includegraphics[width=\hsize]{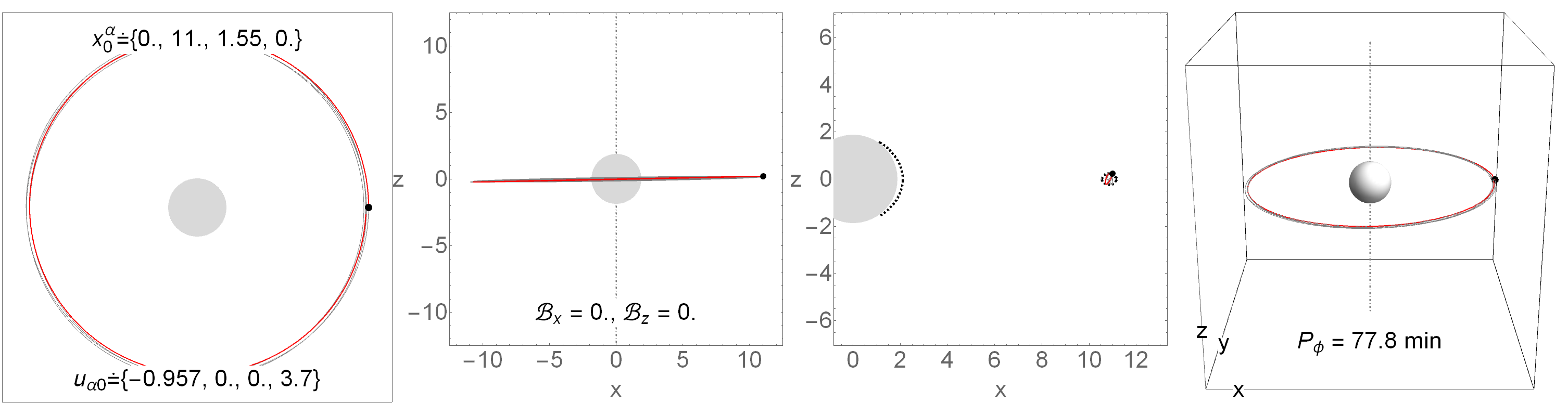}
  \includegraphics[width=\hsize]{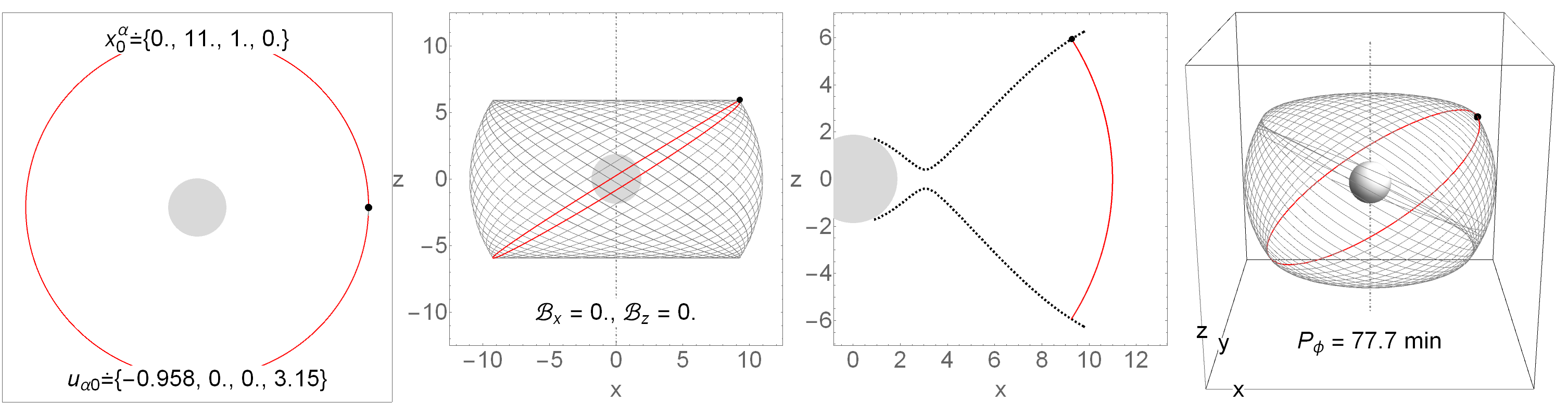}
  \caption{Trajectories of neutral hot spots orbiting Sgr~A* (electromagnetic interaction is neglected) for various relative inclinations of orbital plane and the black hole spin. The first two rows correspond to the starting positions of hot spots at a distance of  $6R_{\rm g}$, while the last two rows are plotted for hot spots starting at a distance of $11R_{\rm g}$ from the black hole. The first column represents a face-on view of the orbit; second column shows the trajectories viewed from the equatorial plane, where the dot--dashed line shows the axis of black hole spin. The third column shows the cross section of the trajectory in a plane orthogonal to the equatorial plane,   and the fourth column represents the 3D trajectories of hot spots with the orbital periods found numerically. See the detailed discussion in Section \ref{sec:sub35}.
}
  \label{fig4}
\end{figure*}

It is interesting to discuss the possibility of misalignment of the orbital planes of the flare components with respect to the rotation axis of the black hole, as the direction of the spin axis of the black hole at the Galactic center remains unknown. Recent studies based on 3D (magneto)hydrodynamic simulations have evidenced a tendency toward alignment of accretion disk structures with the black hole spin in the region close to the black hole, while at large distances, the accretion disk remains at its initial arbitrary orientation \citep{2019arXiv190408428L,2019ApJ...878..149H,2015MNRAS.448.1526N}. The effect is caused by torque that is relatively strong around the black hole and known as the Bardeen--Petterson effect \citep{1975ApJ...195L..65B}. In the case of recent hot spots detected close to Sgr~A*, the rotation of the polarization angles of the synchrotron emission from the positions of flares with the periods approximately equal to the orbital periods of the flare components are consistent with the magnetic field lines orthogonal to the orbital plane \citep{2018A&A...618L..10G}, although the direction of the spin of the black hole remains unknown. Since the orbital axis of the hot spots is  nearly orthogonal to the Galactic plane, it is especially interesting whether the Bardeen-Petterson effect may take place in the case of the Galactic center \citep{2013MNRAS.432.2252D}. For the analysis of hot spot dynamics, though, we will use a charged test particle model.  

A solution of Maxwell's equations describing an asymptotically uniform magnetic field with an arbitrary inclination angle with respect to the spin axis was found by \cite{1985MNRAS.212..899B} that also contains a possible effect of the black hole charge. Introducing two components of a magnetic field $(B_x, B_z)$ with the axis of rotation coinciding with the $z$-axis, one can write the components of the four-vector potential in the following form \citep{1985MNRAS.212..899B,2014ApJ...787..117K}: 
\begin{eqnarray} \label{vecpotbicak}
A_{t}&=&\frac{B_{x}\, a\, M\sin2\theta}{2\Sigma}\left(r\cos\psi-a\sin\psi\right)\\& &\nonumber+ \frac{B_{z}aMr}{\Sigma}\left(1+\cos^2\theta\right)-B_{z}a ,
\\ 
A_{r}&=&- \frac{1}{2} B_x(r-M)\sin2\theta\sin\psi , \\
A_{\theta}&=&-B_xa(r\sin^2\theta+M\cos^2\theta)\cos\psi\\& &\nonumber-B_x(r^2\cos^2\theta-Mr\cos2\theta+a^2\cos2\theta)\sin\psi , \\
A_{\varphi}&=&B_z\sin^2\theta\left[\frac{1}{2}(r^2+a^2)-\frac{a^2Mr}{\Sigma}(1+\cos^2\theta)\right]\label{vecpotbicak2}\\& &\nonumber-B_x\sin\theta\cos\theta\Big[\Delta\cos\psi\\& &\nonumber+\frac{(r^2+a^2)M}{\Sigma}\left(r\cos\psi-a\sin\psi\right)\Big],
\end{eqnarray}
where $\Sigma = r^2+a^2\cos^2\theta$, $\Delta = r^2 - 2 M r + a^2$ and $\psi$ is defined as
\begin{equation}
\label{kicpsi}
\psi=\varphi+\frac{a}{2\sqrt{M^2-a^2}}\ln{\frac{r-M - \sqrt{M^2-a^2}}{r-M + \sqrt{M^2-a^2}}},
\end{equation}
asymptotically approaching $\varphi$, i.e. $\lim_{r\to \infty}\psi=\varphi$.  Here we use the geometrized units, in which $G=c=1$, so that $M$ has the unit of length ($M\rightarrow G M/c^2$). 

In the presence of a magnetic field, the dynamical equations for the motion of charged particles (hot spots) around a black hole are not separable, which can lead to the chaotic character of the motion. Various features of the dynamics of charged particles in combined strong gravitational and magnetic fields have been studied recently (see, e.g. \cite{2020Univ....6...26S,2019EPJC...79..479P,2016PhRvD..93h4012T,2016EPJC...76...32S,2015CQGra..32p5009K,2014ApJ...787..117K}). 
The motion is always regular if the trajectories of the particles are bounded in the vicinity of the equatorial plane, which corresponds to the motion close to the local minimum of the effective potential with circular or quasi-harmonic oscillatory motion of the charged particles. This type of motion modeled for Keplerian accretion disks of stellar mass black holes has been successfully applied for the explanation of the quasi-periodic oscillations of X-ray flux from several microquasars \citep{2017EPJC...77..860K,2018PAN....81..279T}.

However, the inclination of the angle of the hot spot orbit from the equatorial plane or of the magnetic field lines with respect to the axis of black hole rotation may lead to the occurrence of  chaotic behavior of the motion due to breaking the axial symmetry of the system. In that case, the angular momentum of a hot spot is not conserved along its trajectory, which means that the orbital period cannot be properly defined. In particular cases studied below, the hot spot may exhibit quasi-circular motion along a single orbit; after that, the quasi-circular character of the motion breaks and the motion becomes chaotic. 

We numerically solve the equations of motion for a charged hot spot, applying  the external magnetic field given by the solution (\ref{vecpotbicak})-(\ref{vecpotbicak2}). 
Aligning the orbital and magnetic field axes, we investigate the effect of their inclinations with respect to the axis of the black hole rotation. The relative influence of magnetic and gravitational forces is given by the dimensionless parameter ${\cal B} = 0.01$ that corresponds to the hot spot with a small net charge density of the order of $\rho_q \sim 10^{-5}$cm$^{-3}$ if the total number density $\rho_N \sim 10^{6}$cm$^{-3}$, and $\rho_q \sim 10^{-1}$cm$^{-3}$ if $\rho_N \sim 10^{8}$cm$^{-3}$, as derived in Section~\ref{sec:sub32}. In all figures, the starting position of the hot spots is set to $r_0 = 6 R_{\rm g}$ and the black hole spin is aligned in the vertical direction with a fixed value of  $a=0.5M$.  We denote the inclination angle between the black hole spin and initial orbital position of the hot spot as by $\theta_0$. Based on the GRAVITY measurements of the rotation of the polarization plane at the positions of the hot spots, one can assume that the magnetic field is oriented approximately perpendicular to the orbital plane. Therefore, one can introduce a ``more fundamental" angle of inclination between magnetic field and the spin axis, being $\alpha = 90^\circ - \theta_0$.   

In Figure~\ref{fig3} we plot the trajectories of charged hot spots orbiting SMBH Sgr~A* for various relative inclinations of magnetic field lines and the black hole spin.  The first column represents a face-on view of the orbit; i.e., the orbital plane is perpendicular to the line of sight. Initial position and velocity components are given inside the plots in dimensionless units. The second column represents the trajectories viewed from the equatorial plane, with the axis of black hole rotation shown as a dot--dashed line.  Magnetic field  components inclined with respect to the spin axis of the black hole coinciding with z-axis are indicated in the plot. The inclination angle between the magnetic field axis and the black hole spin is given by 
\begin{equation}
    \alpha = \arctan{\left|\frac{B_x}{B_z}\right|} \times \frac{180^\circ}{\pi} \equiv 90^\circ - \theta_0,
\end{equation} 
where $\theta_0$ is an angle between the initial orbital position of the hot spot and the spin of the black hole. Equatorial motion corresponds to $\theta_0 = 90^{\circ}$, or $\alpha = 0^\circ$.   
The third column shows the cross section of the trajectory in a plane orthogonal to the equatorial plane of a black hole, and the last column represents the 3D trajectories of the hot spots with the orbital periods found numerically using a Fourier transform applied to the particles' trajectories.

\begin{figure*}
  %\centering
  \includegraphics[width=\hsize]{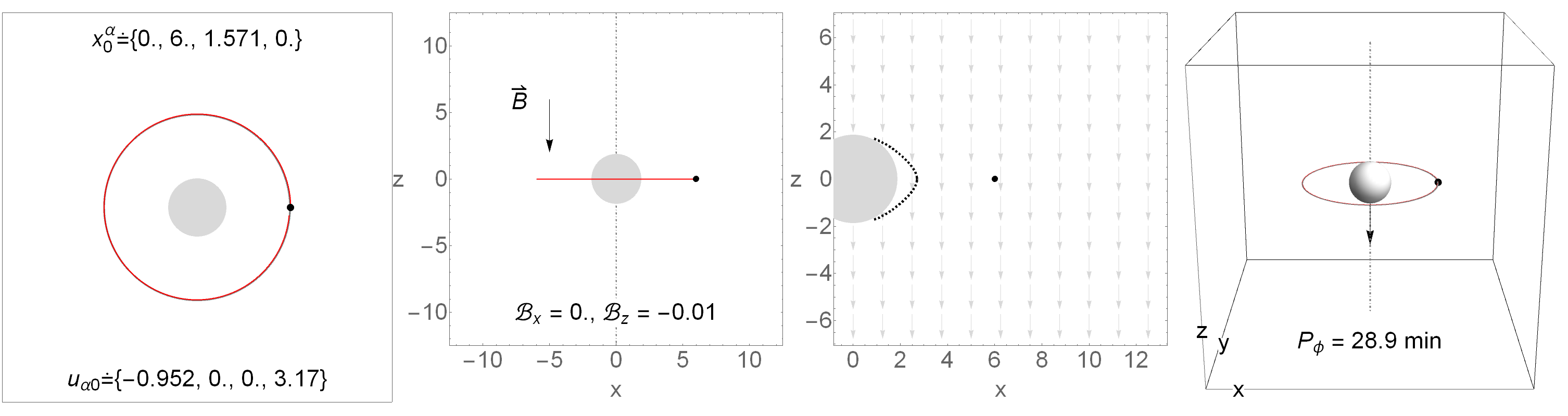}
  \includegraphics[width=\hsize]{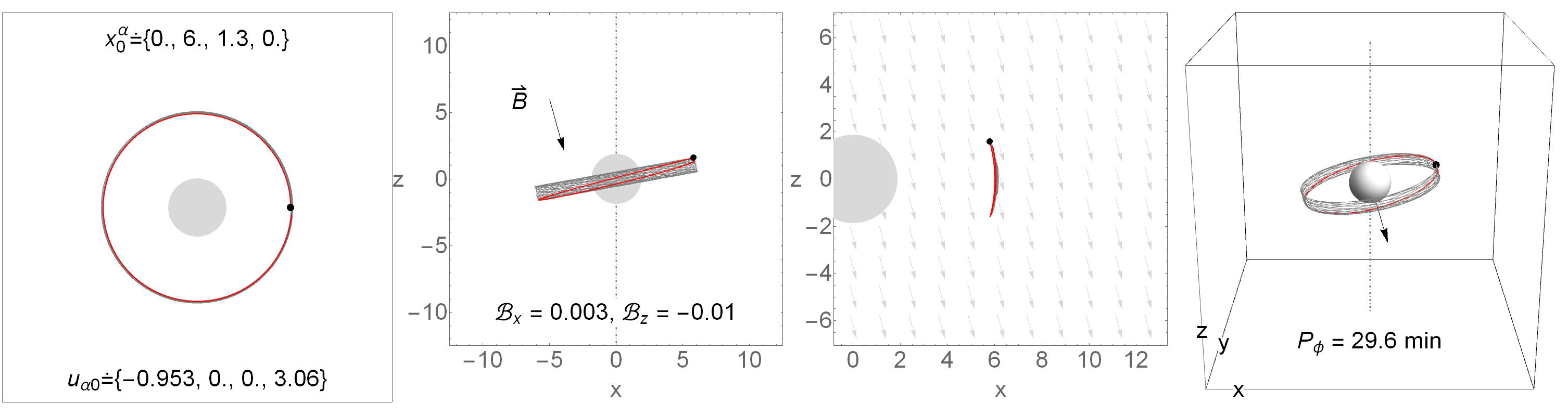}
  \includegraphics[width=\hsize]{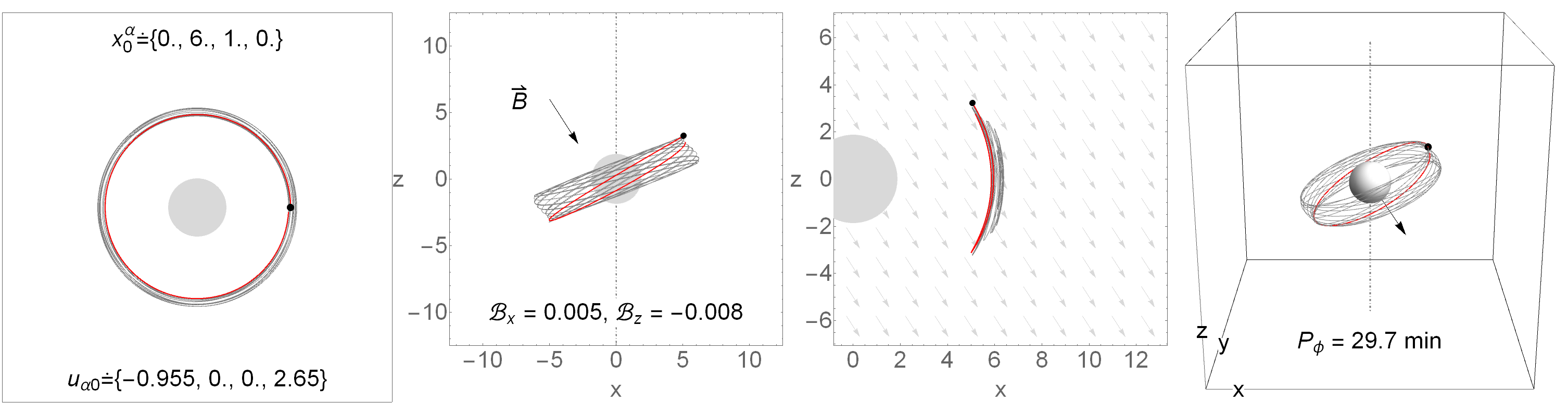}
  \includegraphics[width=\hsize]{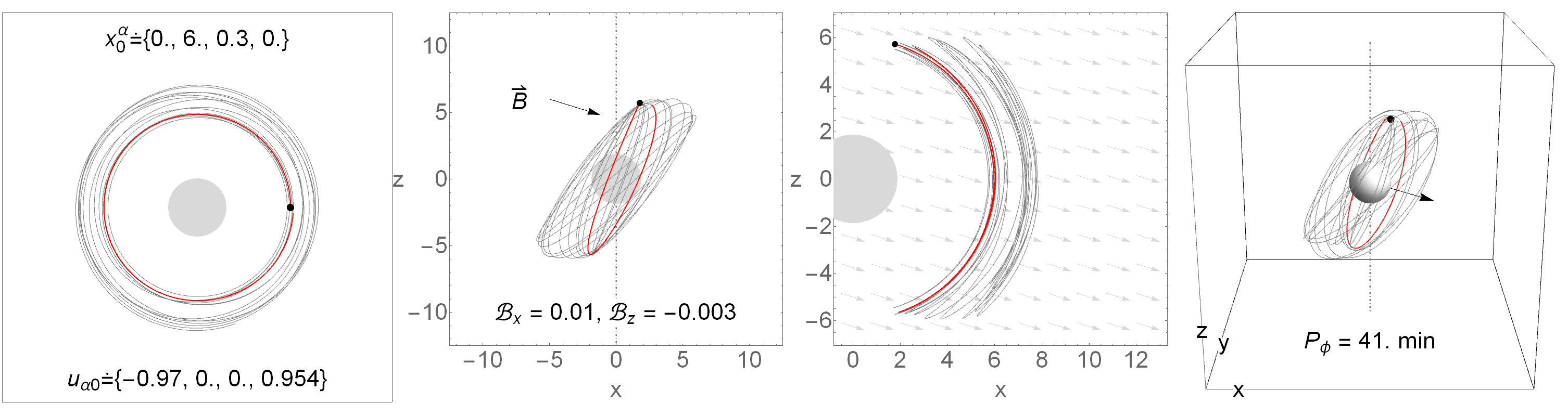}
  \caption{Trajectories of charged hot spots orbiting the SMBH Sgr~A* in the case of opposite orientation of magnetic field axes and its inclinations with respect to the black hole spin that is vertically directed in all plots. See, for comparison, Figure~\ref{fig3} and discussion given in Section~\ref{sec:sub35}.
}
  \label{fig5}
\end{figure*}

 Increasing the relative inclinations of the magnetic field and spin axes leads to the growth of the orbital period of the hot spots for the same orbital positions, while the stability of the orbit decreases. However, for large inclinations, when the spin and magnetic field axes are nearly orthogonal, the motion becomes chaotic, and the orbital period cannot be defined. Therefore, these cases can be excluded according to our model.  From numerical simulations, we find that the stable orbits exist until the critical angle is reached, i.e., in the range of angles, $ 40^{\circ} \lesssim \theta_0 \lesssim 140^{\circ}$, or, equivalently, $|\alpha| \lesssim 50^{\circ}$. This implies that the magnetic field lines have a tendency to be aligned with the black hole spin. 
 Since the trajectories of the closest flare components have been observed with a nearly face-on circular shape, following the above analysis, it seems more likely that the black hole Sgr~A* is aligned toward the observer, rather than perpendicular to the Galactic plane. The Above results derived for the spin $a=0.5M$ are also applicable for arbitrary values of the spin $a > 0.5M$, since the stability of the motion of the charged hot spot at an inclined orbit decreases with increasing the spin of the black hole.

For completeness of results, we performed similar analyses in the case of the neutral hot spot. Results are presented in Figure \ref{fig4}, where the effect of the electromagnetic interaction is neglected. In this case, the equations of motion are fully integrable, and the motion is regular. As opposed to the charged case, the angular momentum is conserved for arbitrary orbital inclinations with respect to the spin that gives a rise to the existence of the boundaries of the motion, plotted in the third column. The first two rows correspond to the hot spot orbiting at a distance of $6 R_{\rm g}$ from the black hole, while the last two rows correspond to $11 R_{\rm g}$. In the neutral case, the orbital periods of hot spots is nearly independent from the orbital inclinations. 

Another possibility of a special interest is the case where the spin and magnetic field axes are oriented in opposite directions. Although such a case seems less likely, one cannot entirely exclude it, as demonstrated in Figure~\ref{fig5}. In contrast to the case represented in Figure~\ref{fig3}, one can also find circular orbits for inclination angles $\theta_0 < 40^{\circ}$;  however, these orbits are highly unstable.

One of the interesting continuations could be the fitting of the detected hot spot orbits with the off-equatorial orbits of \cite{2008CQGra..25i5011K,2010CQGra..27m5006K} that can be stable in the presence of a magnetic field for charged hot spots.

\section{Discussion} \label{sec:discussion}

In this study, we focused on the effect of the ordered, poloidal magnetic field on the plasma components associated with NIR/X-ray flaring activity of Sgr~A*. The motion of plasma within this field leads to a small charge density of only $10^{-3}-10^{-4}\,{\rm cm^{-3}}$, which can, however, have considerable effects on the component motion in this magnetic field. 

\subsection{Flare Components and Their Relation to Large-scale Structures}

We focused on the observed motion of three hot spots observed by GRAVITY \citep{2018A&A...618L..10G} that exhibit  clockwise motion with inclinations $i=160^\circ \pm 10^\circ$. The polarization rotation implies a magnetic field oriented approximately parallel to the vector of the orbital plane, i.e., a poloidal configuration. On the other hand, the bright X-ray flares often exhibit a double-peak structure \citep{2017MNRAS.472.4422K,2017MNRAS.468.2447P}, which is fitted well by the combination of lensing and Doppler boosting of  hot spot emission due to its motion along the trajectory that is close to edge-on with respect to the observer \citep{2018arXiv180600284E}. Already the first hot spot models of Sgr A* flaring activity implied the departure from face-on orientation, with an inclination of $i\leq 145^\circ$ \citep{2006A&A...460...15M}. Hence, the total range of inclinations is expected to be large, $i=90^\circ-160^\circ$, i.e., from nearly edge-on to nearly face-on orbits. Future GRAVITY observations will likely provide better statistics to assess the inclination distribution. 

A large range of hot spot inclinations is generally expected in hot accretion flows \citep{2014ARA&A..52..529Y}, whose half-thickness scales approximately as $h\sim r$; i.e., the half-opening angle of the hot flow is close to $45^\circ$. The mid-plane of the flow is currently uncertain. Taking into account the inflow--outflow accretion flow model, in which most of the material is provided by stellar winds of about 200 OB Wolf--Rayet stars in the central parsec with the mass-loss rates of $\dot{M}_{\rm w}\approx 10^{-5}\,M_{\odot}\,{\rm yr^{-1}}$ and wind velocities of $\sim 1000\,{\rm km\,s^{-1}}$ \citep{2008MNRAS.383..458C,2010ApJ...716..504S,2018MNRAS.478.3544R,2019arXiv191006976C}, the accretion flow midplane could be associated with the clockwise disc of massive OB stars \citep{2003ApJ...590L..33L}, whose inclination is $i=122^\circ \pm 7^\circ$ \citep{2010RvMP...82.3121G}. 
Magnetohydrodynamic simulations of the inner accretion flow of Sgr ~A* that is supplied by stellar winds confirm a broad range of angular momenta and a large thickness of the hot flow \citep{2018MNRAS.478.3544R}. Therefore, the frequent occurrence of hot spots on clockwise orbits with various inclinations ranging from edge-on ($\gtrsim  90^{\circ}$) to nearly face-on ($\lesssim 180^{\circ}$) could also be linked to the hot, thick flow that is supplied by the stellar winds of young stars concentrated at inclinations close to $120^\circ$. Recently, \citet{2019Natur.570...83M} found observational evidence for a  presence of the disk-like structure with a radius of $\sim 0.004\,{\rm pc}$ based on the detection of a broad, double-peak $1.3-{\rm mm}$ H30$\alpha$ line with ALMA. This colder disk has a temperature of $\sim 10^4\,{\rm K}$ and a number density of $10^5$-$10^6\,{\rm cm^{-3}}$,   depending on its exact filling factor. Such a disk-like structure embedded within the hot diluted plasma could be a result of the ``disk'' phase of the evolution of a system of hot Wolf--Rayet stars after $\gtrsim 3000\,{\rm yr}$ \citep{2019arXiv191006976C}.  \citet{2008MNRAS.383..458C} calculated the mean value of the circularization radius $R_{\rm circ}\sim 0.05''=0.002\,{\rm pc}$ of the gas that originates in the stellar winds, which is comparable to the radius of the colder disk of \citet{2019Natur.570...83M}. In addition, the statistics of NIR polarization data implies a rather stable geometrical orientation of the system.
\citet{2015A&A...576A..20S} found typical polarization degrees
of the order of $20\% \pm 10\%$ and a preferred polarization angle of $13^{\circ} \pm 15^{\circ}$.
This orientation is to the first order consistent with that of the
the primary He star clockwise disk \citep{2003ApJ...590L..33L}.

Our analysis is generally applicable to the accretion flows where the magnetic field is dynamically significant,  which seems to be the case for the Galactic center, according to our estimate of the magnetization parameter; see Section~\ref{subsec:MF}. This is especially the case for less dense hot flows, where the frozen-in magnetic field is dragged inward by the accreting gas and accumulates at the very center \citep{1974Ap&SS..28...45B,1976Ap&SS..42..401B} with the dominant poloidal component, which at a certain point is strong enough to disrupt the axisymmetric flow at the magnetospheric radius.  In Section~\ref{subsec: flare-hot spot}, we estimated the magnetospheric radius $R_{\rm m}$ for the spherical flow; see Eq.~(\ref{eq_magnetospheric_radius}).  For the axisymmetric flow, inside $R_{\rm m}$, the flow is supported against gravity by the magnetic pressure, $GM\Sigma/R\approx 2 B_{\rm R}B_{\rm z}/4\pi$, where the surface density of the flow depends on the accretion rate $\dot{M}$ and the radial velocity, which is by a factor of $\epsilon \approx 0.01-0.001$ smaller than the freefall velocity $v_{\rm ff}$ \citep{2003PASJ...55L..69N}, as $\Sigma=\dot{M}/(2\pi R \epsilon v_{\rm ff})$. From this, assuming $B_{\rm R}\sim B_{\rm z}=B_{\rm pol}$, the relation for $R_{\rm m}$ is the following: 
\begin{equation}
    R_{\rm m}\approx \left(\frac{\dot{M}\sqrt{GM}}{\epsilon B_{\rm pol}^2} \right)^{2/5}\,.
    \label{eq_magnetospheric_radius_axi}
\end{equation}
Inside $R_{\rm m}$, the accretion flow is broken up and continues to move inward at velocities much slower than the freefall velocity, $v_{\rm R}=\epsilon v_{\rm ff}$. The MAD consists of magnetic islands and magnetic reconnection events that are expected to be frequent enough to give rise to the adiabatically expanding plasmoids or hot spots \citep{2009MNRAS.395.2183Y,2012A&A...537A..52E,2015ApJ...810...19L}, which can explain the observed flares and their multiwavelength properties, mainly the simultaneous NIR/X-ray flares and time-delayed submillimeter/millimeter/radio flares. In fact, the overall picture is consistent for the Galactic center environment including the stellar and gaseous components. If we consider the disk-like component of the accretion flow, which can be refueled by stellar winds \citep{2019arXiv191006976C,2019Natur.570...83M},   with an accretion rate in the inner $100\,R_{\rm g}$ of $\dot{M}\approx 10^{-8}\,{\rm M_{\odot}\,yr^{-1}}$ \citep{2007ApJ...654L..57M} and a poloidal magnetic field of $B_{\rm pol}\approx 10\,{\rm G}$, the estimate of $R_{\rm m}$ follows from Eq.~(\ref{eq_magnetospheric_radius_axi}),

\begin{align}
    R_{\rm m} & \sim 78.5 \left(\frac{\dot{M}}{10^{-8}\,{\rm M_{\odot}\,yr^{-1}}}\right)^{2/5}\left(\frac{M}{4\times 10^6\,M_{\odot}}\right)^{1/5}\times\\\notag 
    & \times \left(\frac{\epsilon}{0.01}\right)^{-2/5} \left(\frac{B_{\rm pol}}{10\,{\rm G}}\right)^{-4/5} R_{\rm g}\,,
\end{align}
which is of a comparable order of magnitude to the value derived for the spherical accretion, see Eq.~(\ref{eq_magnetospheric_radius}). 
So far, the observed flares or hot spots have been located within this radius \citep{2017MNRAS.472.4422K,2018A&A...618L..10G}. The disk plane could also influence the predominant orientation of the flare components as discussed earlier, albeit with a large scatter due to the thickness of the flow. The overall setup of the hot, thick flow with the inner MAD part is illustrated in Fig.~\ref{fig_illustration_MAD}. However, the detailed numerical modeling of stellar wind feeding including the magnetic field is needed to test the self-consistency of this model, which is beyond the scope of this paper. 

The analyses given in this paper could, in general, be applied to the case of the M87 black hole, for which the first black hole image has been obtained by the Event Horizon Telescope (EHT). The EHT image exhibits a bright, short crescent in the SSE--SWW position angle and a relatively compact hot spot at  the SEE sector \citep{2019ApJ...875L...1E}. Fitting the image and the orientation of the jet emission simultaneously with the general relativistic magnetohydrodynamic simulations makes it difficult to explain the emission in the   SEE sector \citep{2019arXiv190810376N}, at least within strictly stationary and axially symmetric models. The presence of the charge in the accretion flow taking into account the electromagnetic interaction of the accretion disk of M87 could potentially help to resolve the problem, which requires further study. We note that the orbital timescale of the flare components in M87 is significantly longer. While for Sgr~A*, we have the basic timescale at the ISCO of $P_{\rm SgrA*}=12\sqrt{6}\pi G M/c^3\sim 31.3\,{\rm min}$, for M87, we have an   orbital timescale of $P_{\rm M87}=(M_{\rm M87}/M_{\rm SgrA*})P_{\rm SgrA*}\approx 1570 \times 31.3\,{\rm minutes}\sim 34\,{\rm days}$. Therefore, if the flare component has a shorter timescale than the orbital timescale, it may be complicated to trace the flare components of M87 along their orbits, as was done for Sgr~A*.   

\subsection{Could Hot Spots Be a Part of an Outflow?}

In our analysis, we considered the orbiting hot spot to arise in the accretion flow. Hence, the orbit is expected to be bounded or inspiralling with respect to the Sgr~A* black hole.  In case the hot spot had a large bulk motion, it could become a part of an outflow or the sheath of a helical jet \citep{2020arXiv200304330R}. Several observations of faint jet-like structures in the vicinity of Sgr A* have been reported in the X-ray, radio, and IR domains, in general at different position angles \citep{2006A&A...455....1E,2012ApJ...758L..11Y,2013ApJ...779..154L,2015A&A...576A..20S}. Although the ADAF-jet model can explain the broadband characteristics of Sgr A* \citep{2002A&A...383..854Y}, there is no clear observational evidence for hot spots to be jet plasmoids on outflowing trajectories. In fact, a significant linear component of the velocity is expected, which for large inclinations of $i\approx 160^{\circ}$ should lead to much stronger Doppler boosting than observed. This can be shown from the general expectation that the jet velocity should be of the order of the local orbital velocity of the accretion material \citep{2005AIPC..784..183D}, $v_{\rm j}\sim v_{\rm orb}(\rm ISCO)\sim 0.5c$. For a small angle to the line of sight, this would lead to a Doppler-boosting factor of $\delta\approx [\gamma(1-\beta)]^{-1}\approx \sqrt{3}$, and the overall Doppler term 
\begin{equation}
    \frac{S_{\rm obs}}{S_{\rm HS}}=D\equiv \delta^{3+\alpha}\approx 7.22-11.84\,,
    \label{eq_doppler_flux}
\end{equation}
where $S_{\rm obs}$ and $S_{\rm HS}$ are the flux densities measured in the observer's frame and the frame of the hot spot, respectively. The spectral index $\alpha$, $S_{\nu}\propto \nu^{-\alpha}$, is 0.6 and 1.5 for the very bright and average bright flares, respectively \citep{2018ApJ...863...15W}. This is clearly in contradiction to the observational constraints on the Doppler term of $D<2.5$. In case the jet components had a significant $\phi$-component of their velocity,   the Doppler term would limit the forward speed to $0.20-0.24\,c$, which is comparable to the projection of a $\phi$-component to the line of sight, $v_{\phi}\sin{i}=0.5\times \sin{20^{\circ}}\approx 0.17$. Hence, the hot spots would be nearly stationary components of the jet. Although in some cases nearly stationary jet components are detected, e.g. in blazar jets \citep[see, e.g.][]{2018MNRAS.478.3199B}, their long-term kinematics evolves on a timescale of years, in comparison with the short-term hot spot clockwise motion as detected by the GRAVITY instrument \citep{2018A&A...618L..10G}.  Therefore, the assumption that hot spots arise in the accretion inflow rather than outflow is better justified, at least for a current set of orbiting hot spots.

\subsection{Comparison with other studies}

 \citet{2020A&A...635A.143G} analyzed the astrometric positions of the three hot spots detected in 2018 with a general relativistic ray-tracing code. They included the effects of out-of-plane orbits, as well as the shearing of hot spots. They inferred a mean orbital radius of $\sim 9\,R_{\rm g}$, and an  inclination of $i\sim 140^{\circ}$, and they constrained the hot spot diameter to less than  $5 R_{\rm g}$, i.e. the hot spots must be very compact emission sites.

 \citet{2020arXiv200304330R} studied the magnetic reconnection and the plasmoid formation in current sheaths in the accretion flows using general relativistic MHD simulations. They confirmed that plasmoids can form in the inner accretion flow parts between 5 and 10 Schwarzschild radii, regardless of the disk size and its magnetization. Plasmoids can further merge and grow to a macroscopic size of the order of a Schwarzschild radius ($\sim 0.1\,{\rm AU}$) and get advected toward the black hole or become a part of the jet sheath.

 \citet{2020arXiv200413029M} attempted to explain an apparent paradox in GRAVITY observations: the hot spots appear to traverse a loop on the sky in a time shorter than expected, which leads to a velocity of $\sim 0.3c$. The authors proposed that instead of material motions, in which the hot spot follows a gas parcel in the RIAF or along a geodesic, the radius--velocity discrepancy can be solved by the pattern motion at super-Keplerian speeds at larger radii. First, they proposed a magnetohydrodynamic perturbation at $r\sim 12.5\,R_{\rm g}$, or, alternatively, the pattern could be created due to the interaction between the outflow and the inclined disk at a radius of $\sim 20\,R_{\rm g}$. 

Our work is complementary to the above-mentioned studies in a way that it considers the model of a charged magnetosphere. In addition, the paradox of super-Keplerian speeds can be solved by the presence of the Lorentz force.

\begin{table*} 
  \centering
  \caption{Summary of the constraints on the parameters of the Galactic center flare components.  \\ Here $\rho_N$ denotes the total number density of particles in the plasma, $\rho_q$ denotes the number density of extra charged particles, $q_{\rm hs}$ denotes the net charge of the hot spot for various methods of estimation, $\alpha$ denotes the angle between the black hole spin axis and the orientation of the magnetic field lines.}
%  \resizebox{\textwidth}{!}{  
  \begin{tabular}{c|c|c|c}
  \hline
  \hline
     Parameter & Limit & Note & Reference\\
  \hline
    $R_{\rm hs}$ & $(1-5) R_{\rm g} { G M}/{c^2} \approx  \, 6\times 10^{11} {\rm cm} -  3 \times 10^{12}   {\rm cm}$ & size & Eqs.~(\ref{Rhs1}), (\ref{Rhs2})\\
     $m_{\rm hs}$ & $m_{\rm hs}^{\rm min}\approx 1.2\times 10^{17}\,{\rm g}$,  $m_{\rm hs}^{\rm max}\approx 2.3\times 10^{20}\,{\rm g}$    & mass & Eq.~(\ref{masshs1})  \\
     $\rho_{\rm N}$ & $\approx 10^{7\pm1} {\rm cm}^{-3}$    & total par. density & Eqs.~(\ref{rhoN1}), (\ref{rhoN2})  \\
     \hline 
    $\rho_q^{\rm mag.}$  & $\approx 1.28 \times 10^{-4} \left(\frac{B}{10{\rm G}} \right) \left( \frac{T}{45 \rm{min}}\right)^{-1} {\rm cm^{-3}}$  & charge separation & Eq.~(\ref{rho-gj-sr}) \\
      $\rho_q^{\rm dyn.}$ & $< 10^{-2} \left(\frac{B}{10 {\rm G}}\right)^{-1} 
 \left( \frac{\rho_{\rm N}}{ 10^{6} \, {\rm cm^{-3}}}  \right)   {\rm cm^{-3}}$  & orbital fitting & Eq.~(\ref{rho-dyn})\\
   $\rho_q^{\rm centroid }$ & $\approx 3 \times 10^{-3} 
     \left(\frac{B}{10 {\rm G}}\right)^{-1} 
 \left( \frac{\rho_{\rm N}}{ 10^{6} \, {\rm cm^{-3}}}  \right)   {\rm cm^{-3}}$ & centroid orb. fit. & Eq.(\ref{rho-mean}) \\
   \hline 
    $|q_{\rm hs}^{\rm mag. }|$ & $2 \times 10^{13} \left(\frac{B}{10{\rm G}} \right) \left(\frac{R}{R_s}\right)^3  {\rm C}$ & charge separation & Eq.(\ref{qGJ1}) \\
    $|q_{\rm hs}^{\rm dyn. }|$ & $< 10^{14.5 \pm 2.5} \left(\frac{B}{10 {\rm G}}\right)^{-1 } 
 \left(\frac{m_{\rm hs}}{10^{17.5 \pm 2.5} {\rm g}}\right) {\rm C}$ & orbital fitting & Eq.(\ref{qhsdyn1}) \\
    $q_{\rm hs}^{\rm centroid }$ & $- 10^{13.5 \pm 2.5} \left(\frac{B}{10 {\rm G}}\right)^{-1 } 
 \left(\frac{m_{\rm hs}}{10^{17.5 \pm 2.5} {\rm g}}\right) {\rm C}$ & centroid orb. fit. & Eq.(\ref{qhsmean1}) \\
    $|q_{\rm hs}^{\rm Coulomb }|$ & $<  10^{15 \pm 3} \left(\frac{Q}{Q_{\rm W}}\right)
  \left(\frac{m}{m_{\rm hs}}\right)$ C & charged BH orb. fit. & Eqs.(\ref{qhs-QQ1}), (\ref{qhs-QQ2}) \\
     $|q_{\rm hs}^{\rm syn. }|$ & $< 10^{14.5 \pm 1.5} \left(\frac{B}{10 {\rm G}}\right)^{-\frac{1}{2} } 
 \left(\frac{m}{m_{\rm hs}}\right)^{\frac{1}{2} }
 \left(\frac{L}{10^{33} {\rm erg \, s^{-1}}}\right)^{\frac{1}{4} } {\rm C}$ & synchrotron fit. & Eqs.(\ref{qhs-synch1}), (\ref{qhs-synch1}) \\
     \hline
      $Q_{\rm Wald}^{\rm max }$ (Sgr A*) & $ \lesssim 10^{15} \left( \frac{M}{4 \times 10^6 M_{\odot}} \right)^2  \left( \frac{B_{\rm ext}}{10 \rm G} \right)~  {\rm C}$  & induced BH charge & Eq.(\ref{eq_charge_wald}) \\
     \hline
        $\alpha$ & $|\alpha| < 50^{\circ}$, for  $a \geq 0.5$  & angle between $a \& B$ & Section~\ref{sec:sub35} \\
         $\theta_0$ & $ 40^{\circ} < \theta_0 < 140^{\circ}$, for  $a \geq 0.5$  & angle between $a ~\&$ orb. plane & Section~\ref{sec:sub35} \\
     \hline
  \end{tabular}
 % }
  \label{tab_summary_constraints}
\end{table*}

\section{Summary and Conclusion} \label{sec:conclusion}

The main results of the paper related to the constraints on some parameters of the flares and their components are summarized in Table~\ref{tab_summary_constraints}.

 Measurements of the rotations of the polarization planes at the Galactic center suggest the existence of the large-scale, ordered magnetic field in the close vicinity of the SMBH  Sgr~A*. 
Assuming the presence of the orthogonal component of the magnetic field, the motion of relativistic plasma around the black hole leads to the charge separation and resulting nonnegligible net charge density in a plasma. On the other hand, the rotation of the black hole in the magnetic field induces the electric field on the black hole surface. As a result, both the black hole and its magnetosphere possess nonzero and opposite electric charges that are gravitationally weak but electromagnetically nonnegligible.

In black hole case the charge is given by the solution of  \cite{1974PhRvD..10.1680W}, for the charge of black hole magnetosphere by \cite{1975PhRvD..12.2959R}. A special relativistic analog of the described charging mechanism and applied to the rotating magnetized neutron stars is known by the solution of \cite{1969ApJ...157..869G}.

Applying a simplified toy model of the axially symmetric electron--proton plasma hot spot with a magnetic field strongly orthogonal to the hot spot's orbital plane, we parameterized the ratio of the Lorenz force to the gravitational force  by the dimensionless parameter ${\cal B}$, given by Eq.~(\ref{BBparam}) and constrained its value in between $10^{-5}$ and $10$. In general, this implies that the electromagnetic forces acting on the hot spots moving around Sgr~A* may dominate over the gravitational force from the SMBH.

Tighter constraints on the parameter ${\cal B}$ were made by analyzing the dynamics of the components of three recent flares detected by the \cite{2018A&A...618L..10G} near the ISCO of Sgr~A*. Note that the positions of the centroids of all three hot spots are slightly lower in periods than expected for the Kerr black hole case (in the absence of electromagnetic interaction). As we have shown in Figure~\ref{fig1}, the centroids can be well fitted if the electromagnetic interactions are taken into account with the parameter $\BB \approx  - 3 \times 10^{-3}$. This corresponds to a net charge number density of the order of $3 \times 10^{-3}$cm$^{-3}$, if the total number density of particles in the accretion flow is of the order of $10^{6}$cm$^{-3}$. It is important to note that centroid fittings suggest that the sign of the hot spot is negative, while the black hole charge, in general, is more likely to be positive. One can also see that this result is in accordance with \cite{1975PhRvD..12.2959R} charging mechanism, where the charge of the plasma magnetosphere is equal with opposite sign to the charge of the black hole. 
Note that these constraints are applicable, in general, to an arbitrary hot spot model, constituted by plasma.  

One of the most important consequences of the electromagnetic interaction is the shift of the location of the ISCO of charged hot spots with respect to the neutral case. The results of the ISCO shifts in a simplified case of a uniform magnetic field are shown in Figure~\ref{figisco}. The value of the ISCO shift depends on the sign of the hot spot charge among other parameters. For a positive (negative) hot spot charge and magnetic field aligned along the orbital axis, the electromagnetic parameter ${\cal B}$ can mimic the spin of the black hole up to a value of $a=0.93$ (or $a=0.48$ in the negative case). Since the location of the ISCO is among the most important parameters determining the black hole spin in the accretion theory, a possible discrepancy in the measurements of the spin due to the magnetic field has to  be properly analyzed.  

Comparing the observed luminosity of the flares with those of the charged hot spot, we obtained another constraint on the charge of the flare components in Section~\ref{sec:synchrotron} that is of a similar order of magnitude as constraints from dynamics. In order to complete the study, we also considered  purely Coulombic interactions between the black hole (with the Wald charge) and the hot spots that led us to another constraint, given in Section~\ref{sec:bhcharge-dynamics}. We  also estimated the size and mass of the hot spots in Section~\ref{sizeandmass}.

We have shown that the dynamics of the matter at the Galactic center can be chaotic if the axial symmetry of the system is broken. Analyzing the motion of hot spots around a black hole in a magnetic field inclined with respect to the black hole spin and comparing the trajectories with those of the observed flare components,  we have constrained the inclination angle between the black hole spin and the axis of the hot spot orbital plane to $\alpha < 50^{\circ}$. For larger angles, the motion is strongly chaotic, and the orbital frequency cannot be properly determined.  This, in particular, excludes the case of the orientation of the black hole with the Galactic rotation axis. On the other hand, our constraints on the orientation of the black hole spin make sense only for relatively rapid spins with $a>0.5$. If the black hole is rotating slowly or not rotating at all, the orbital periods, positions, and shapes of the trajectories can be fitted for a black hole with arbitrary spin inclinations.

The importance of electromagnetic interactions of flare components with external electromagnetic fields lies in the consequently arising  uncertainty in the determination of the black hole spin. Depending on the orientation of the Lorentz force and the electrostatic interaction between the hot spot and black hole, the orbital period and radius of the hot spot can be shifted in both ways. This can mimic the effect of the black hole spin on the position of the ISCO. Moreover, the charging mechanisms are crucial for understanding of the energy extraction processes from black holes, as the induced electric field is the driving force in the jet acceleration models based on, e.g.,  \cite{1977MNRAS.179..433B} and magnetic Penrose process \citep[see, e.g. recent review by][]{2019Univ....5..125T}. Therefore, the investigation of the electromagnetic effects in the dynamical environment of the Galactic center,  requires further, more sophisticated analyses and simulations.

\acknowledgments

A.T. was supported by the International Mobility Project CZ.02.2.69/0.0/0.0/16\_027/0008521. 
M.Z. and B.C. acknowledge the financial support by the National Science Centre, Poland, grant No. 2017/26/A/ST9/00756 (Maestro 9). 
A.E. and V.K. are grateful for support through the
German Research Foundation (DFG) program
E~137/10-1 and the Czech Science Foundation grant No. 19-01137J.
Z.S. acknowledges the Czech Science Foundation grant No. 19-03950S.
 This project was also carried out within the
Collaborative Research Centre 956, sub-project [A02],
funded by the Deutsche Forschungsgemeinschaft (DFG) project, ID 184018867.

\end{document}